\documentclass[a4paper,fleqn,usenatbib]{mnras}

\usepackage[T1]{fontenc}
\usepackage{ae,aecompl}

\usepackage{longtable}
\usepackage{graphicx}	
\usepackage{amsmath}	
\usepackage{amssymb}	

\usepackage{pdflscape}
\usepackage{multicol}

\title[Active phases and flickering of a SyRN T CrB]{ Active phases and flickering of a symbiotic recurrent nova T CrB}
\author[I{\l}kiewicz et al.]{Krystian I{\l}kiewicz,$^{1}$\thanks{E-mail: ilkiewicz@camk.edu.pl} Joanna Miko{\l}ajewska,$^{1}$ Kiril Stoyanov,$^{2}$ \newauthor Antonios Manousakis$^{1}$ and Brent Miszalski$^{3,4}$ \\
$^{1}$Nicolaus Copernicus Astronomical Centre, Bartycka 18, 00716 Warsaw, Poland \\
$^{2}$Institute of Astronomy and National Astronomical Observatory, Bulgarian Academy of Sciences, Tsarighradsko Shose 72, \\BG-1784 Sofia, Bulgaria \\
$^{3}$South African Astronomical Observatory, PO Box 9, Observatory, 7935, South Africa \\
$^{4}$Southern African Large Telescope Foundation, PO Box 9, Observatory, 7935, South Africa }

\date{Accepted XXX. Received YYY; in original form ZZZ}

\pubyear{2015}

\begin{document}
\label{firstpage}
\pagerange{\pageref{firstpage}--\pageref{lastpage}}
\maketitle

\begin{abstract}
   T~CrB is a symbiotic recurrent nova known to exhibit active phases, characterised by apparent increases in the hot component temperature and the appearance of flickering, i.e. changes in the observed flux on the time-scale of minutes. Historical UV observations have ruled out orbital variability as an explanation for flickering and instead suggest flickering is caused by variable mass transfer. We have analysed optical and X--ray observations to investigate the nature of the flickering as well as the active phases in T~CrB. The spectroscopic and photometric observations confirm that the active phases follow two periods of $\sim$1000d and $\sim$5000d.  Flickering in the X--rays is detected and follows an amplitude--flux relationship similar to that observed in the optical. The flickering is most prominent at harder X--ray energies, suggesting that it originates in the boundary layer between the accretion disc and the white dwarf. The X--ray radiation from the boundary layer is then reprocessed by a thick accretion disc or a nebula into UV radiation. A more detailed understanding of flickering would benefit from long-term simultaneous X--ray and optical monitoring of the phenomena in symbiotic recurrent novae and related systems such as Z~And type symbiotic stars.
\end{abstract}





\section{Introduction}

Cataclysmic variables (CVs) are interacting binaries in which a white dwarf (WD) is accreting matter from a late--type donor. In the classical picture of a CV a Roche lobe filling star transfer mass through an accretion disc to the WD. In some CVs the material accreted onto a WD reaches a pressure and temperature sufficient to trigger a thermonuclear reaction, which gives rise to a classical nova outburst. If more than one outburst is observed, then it is classified as a recurrent nova (RN; see e.g. \citealt{2003cvs..book.....W}). If the donor star is also a red giant (RG), it is classified as a symbiotic recurrent nova (SyRN; see e.g. \citealt{1999A&A...344..177A}; \citealt{2011A&A...527A..98S}), with the system also belonging to the family of symbiotic stars (SySt; see \citealt{2012BaltA..21....5M} for a recent review).

T~CrB is a SyRN with two outbursts recorded to date (Nova CrB 1866, 1946). The RG in the system has a spectral type M4III \citep{1999A&AS..137..473M} and fills its Roche Lobe \citep{1998MNRAS.296...77B}. The masses of the components are M$_{\mathrm{WD}}=1.2\pm0.2$M$_\odot$ and M$_{\mathrm{RG}}=0.8\pm0.2$M$_\odot$  \citep{1998MNRAS.296...77B,2004A&A...415..609S}. The average quiescent luminosity of the WD is close to 40~L$_\odot$ \citep{1992ApJ...393..289S}. T CrB was discovered as an X--ray source by \citet{1981ApJ...245..609C} and the X--ray emission has a relatively hard spectrum for a symbiotic star \citep{2008ASPC..401..342L}.

Occasionally between the nova eruptions, T~CrB and other SyRNe (V745~Sco, V3890~Sgr and RS~Oph) show enhanced activity of the hot component, manifested by an increase in the emission line fluxes and appearance of a blue continuum. In particular, the \mbox{He\,{\sc i}} and \mbox{He\,{\sc ii}} emission lines became  clearly visible in the spectrum whereas there are very weak or absent during most of the quiescent phase  \citep{1990JAVSO..19...28I}. The line variability is correlated with the changes in the UV continuum and there is no correlation with the orbital period \citep[e.g.][]{1985ESASP.236..213C}. Similar activity was reported in the episodic appearance of the \mbox{He\,{\sc ii}\,4686} emission line and a hot continuum in the optical spectrum before the 1946 outburst (e.g. \mbox{\citealt{1939ZA.....17..246H}}; \citealt{1943PASP...55..101M}). The nature of these active phases is yet fully understood.

One of the most mysterious phenomena observed in T~CrB is flickering. Flickering is a stochastic variation in the light curve of a star with an amplitude of few tenths of a magnitude on time-scales of seconds or minutes. It is well known that flickering in CVs is associated with accretion onto the WD, but the exact physical process is unknown. One of the first models of flickering included unsteady accretion through a bright spot \citep{1971MNRAS.152..219W}. The first to conduct a systematic study was \citet{1992A&A...266..237B} who estimated theoretical energies and time-scales of flickering. The author considered unstable mass transfer and interaction of matter with the disc edge, dissipation of magnetic loops, turbulence in the accretion disc and unstable accretion in the boundary layer. As a result the most promising model contained unstable mass transfer through a boundary layer and turbulence in the accretion disc (see also \citealt{1993A&A...275..219B}). Different models of flickering include magnetohydrodynamic turbulence transporting the angular momentum outward \citep{1998RvMP...70....1B}, turbulent transport of angular momentum in the whole disc \citep{2010MNRAS.402.2567D}, occasional flare-like events and subsequent avalanche flow in the accretion disk atmospheres \citep{1997ApJ...486..388Y}, discrete flares in the accretion disc \citep{2006Ap&SS.304..291R}, or a more recent model by \citet{2014MNRAS.438.1233S} which is mainly based on a fluctuating accretion disc.

The flickering in T~CrB was observed by a number of authors (see \citealt{2006AcA....56...97G} and references therein). The colours of the flickering source in T~CrB indicate a temperature of $\sim$9000~K, which is significantly lower than the average temperature of flickering sources in classical CVs ($\sim$20000~K; \citealt{2015AN....336..189Z}). Flickering in T~CrB seems to originate in the vicinity of the WD \citep{1998A&A...338..988Z}. The ratio of the amplitude of flickering in T~CrB to the average flux remains constant \citep{2004MNRAS.350.1477Z}, which according to a model by \citet{1993A&A...275..219B}, means that the size of the boundary layer remains roughly constant. This is not necessarily the case in all CVs \citep{1998A&A...332..586F}. Flickering was also observed in the X--ray observations of T~CrB \citep{2008ASPC..401..342L} and in the emission lines \citep{2005PASP..117..268Z}.

In this paper we analyse optical and X--ray observations of T~CrB in order to shed light on the nature of active phases and flickering observed in this system. The collected observations are presented in section~\ref{obs_sec}. In section~\ref{active_phase_section} we present results concerning the active phases of T~CrB. Section~\ref{flickering_section} is dedicated to the optical and X--ray observations of flickering in the system. A brief discussion about the nature of the observed phenomena is presented in section~\ref{discussion_sec} and a summary of the results is given in section~\ref{sumarysec}.

\section{Observations}\label{obs_sec}

\subsection{Spectroscopy}
Spectroscopic observations were obtained from the Astronomical Ring for Access to Spectroscopy database\footnote{http://www.astrosurf.com/aras/} (ARAS). A log of observations is presented in Table~\ref{logspec}. The data include mainly low resolution spectra covering a wide spectral range. Basic spectrophotometric calibration of these spectra was obtained by using a standard star spectrum. This procedure was suitable for obtaining the spectral energy distribution (SED). However, absolute flux calibration could not be obtained this way due to light losses on the slit, variable weather conditions and other effects. Therefore, we scaled the spectra using the known $V$ magnitudes (Sect. \ref{sec:phot}). The emission line intensities were measured by Gaussian profile fitting. When a Gaussian did not fit a line, we measured the local continuum level and integrated all of the signal above that level. The main source of uncertainty was the selection of the local continuum level. The uncertainty was estimated by measuring the same emission lines by different people. This procedure resulted in an estimated precision of 15\% and 30\% in the case of strong and faint lines, respectively. From our analysis we excluded two spectra in the ARAS database. The spectra from 01.04.2012 and 27.04.2015 had unusual SEDs that suggested problems with their flux calibration. The calibrated low resolution spectra, measured equivalent widths and fluxes of emission lines are presented in the Appendix \ref{sect:appendix} and Fig.~\ref{all_fluxes}.

\begin{figure}
\begin{center}
\resizebox{\hsize}{!}{\includegraphics{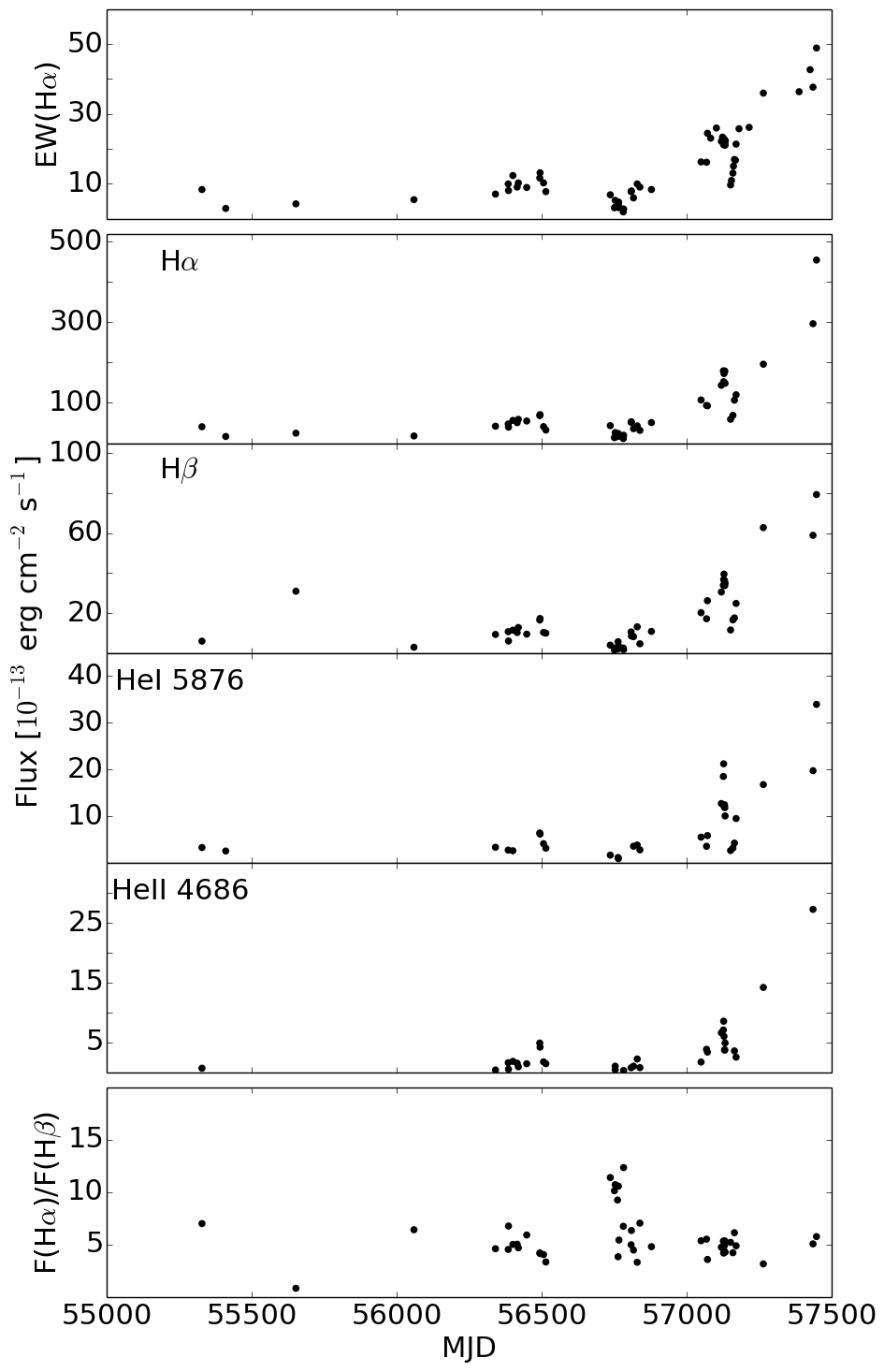}}
\end{center}
\caption{Variability of the equivalent width of H$\alpha$, other selected emission line fluxes and the H$\alpha$ to H$\beta$ flux ratio.}
\label{all_fluxes}
\end{figure}

\subsection{Photometry}
\label{sec:phot}

\begin{figure*}\centering
\resizebox{\hsize}{!}{\includegraphics{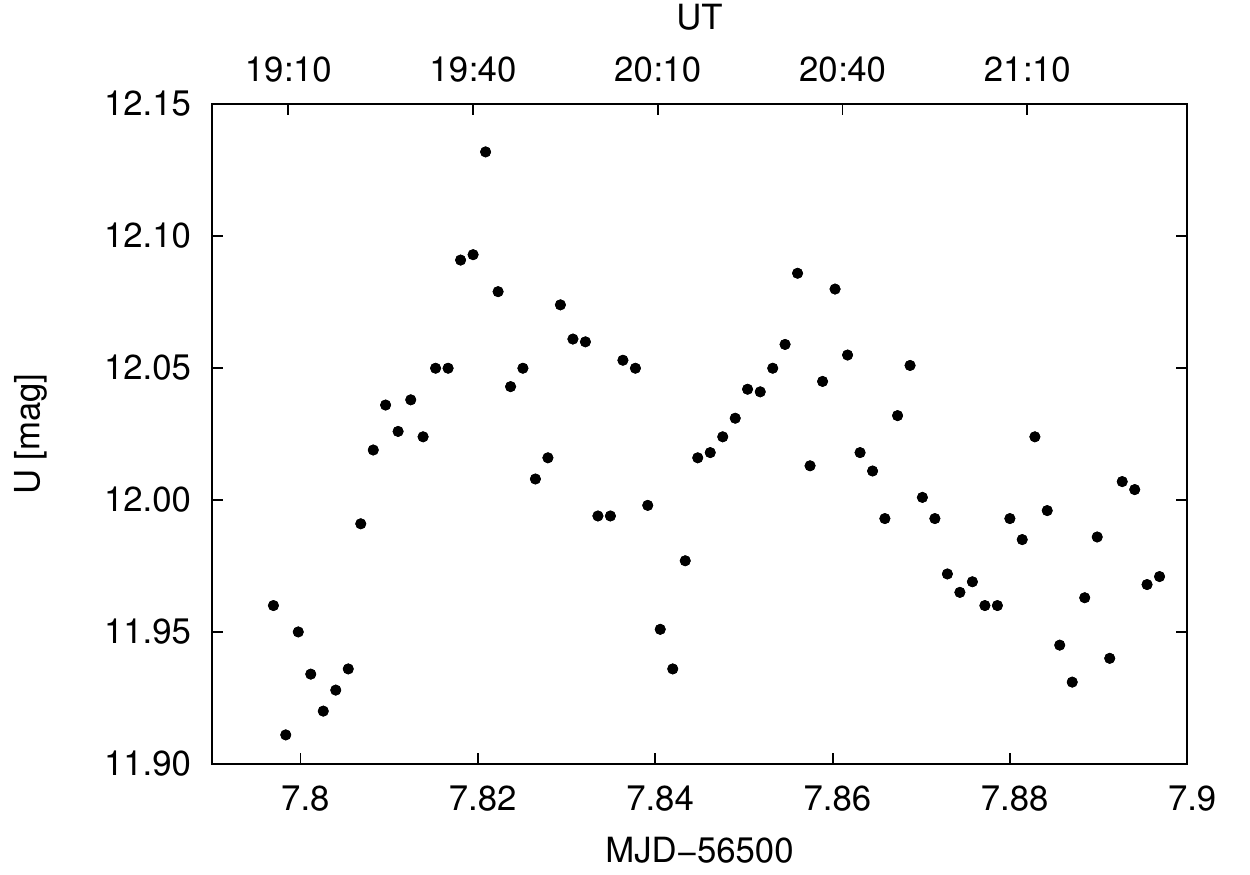}\includegraphics{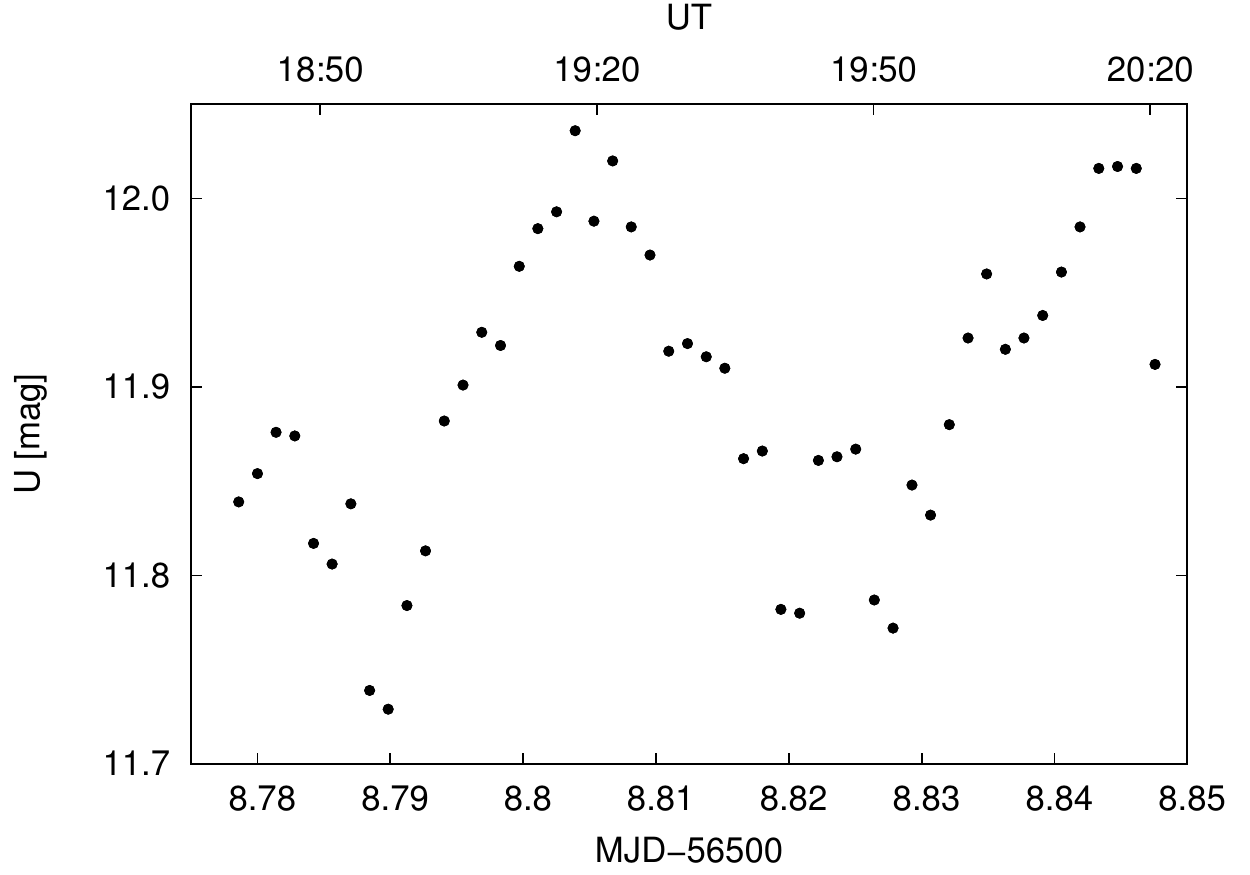}}
\caption{Observations of flickering in T CrB.}
\label{flickering_fig}
\end{figure*}

On the nights of 2013 August 3 and 4, we carried out $U$ band observations of the flickering in T~CrB with the 60cm Cassegrain telescope at Rozhen National Astronomical Observatory, Bulgaria. The telescope was equipped with a FLI~PL09000 CCD camera with 3056x3056 pixels and 27'x27' field of view. Single exposure times of 120~s were used for a total of about 6~hours of observations. Data reduction was carried out using standard IRAF procedures.\footnote{IRAF is distributed by the National Optical Astronomy Observatory, which is operated by the Association of Universities for Research in Astronomy (AURA) under a cooperative agreement with the National Science Foundation.} The typical error of each observation was 0.006~mag. A summary of the observations is presented in Table~\ref{tableflickering}. The light curves of the flickering are presented in Fig.~\ref{flickering_fig}.

\begin{figure*}\centering
\resizebox{\hsize}{!}{\includegraphics{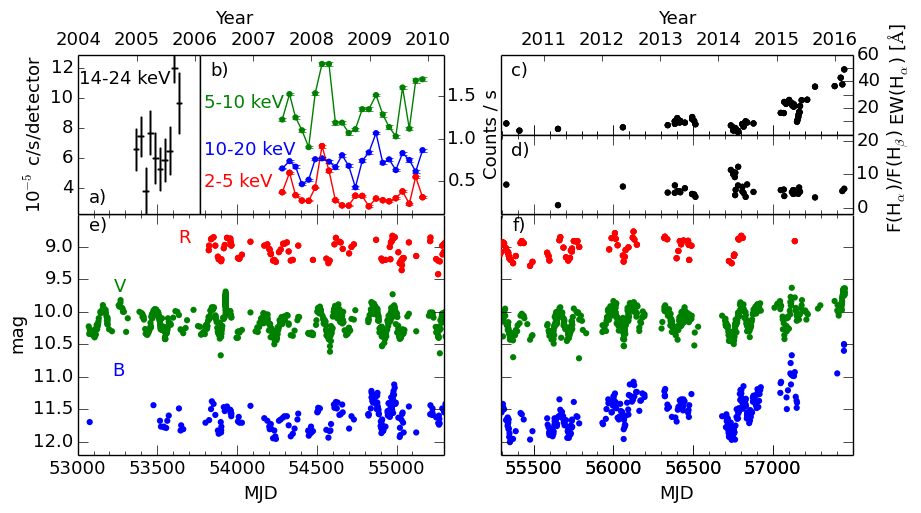}}
\caption{ a) $Swift$ observations of T~CrB from \citet{2009ApJ...701.1992K}; b) X--ray observations from RXTE; c) H$_\alpha$ equivalent width; d) Ratio of H$_\alpha$ to H$_\beta$ flux; e) and f) Photometric observations from AAVSO and ASAS. }
\label{TCrB_all}
\end{figure*}

\begin{figure}\centering
\resizebox{\hsize}{!}{\includegraphics{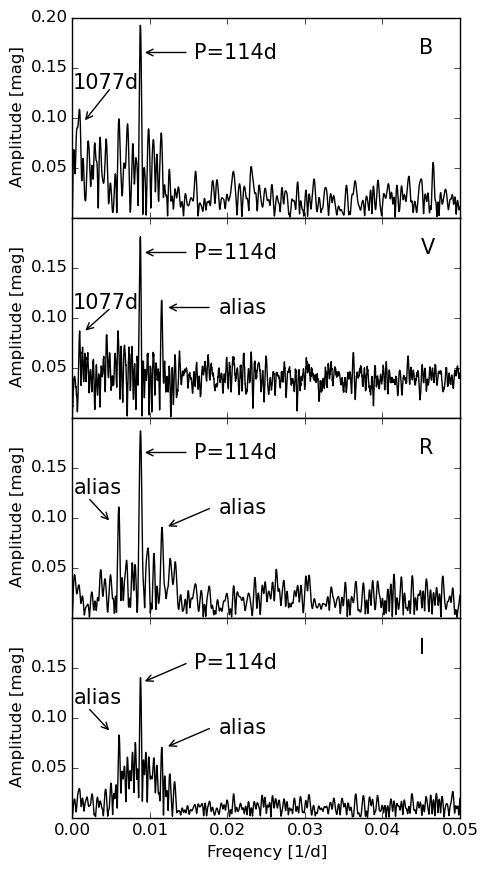}}
  \caption{Power spectra of the AAVSO observations.}
\label{TCrB_powerspec}
\end{figure}

In order to analyse the long term variability we adapted photometry from publicly available catalogues. Namely we used $BVR$ observations from the  American Association of Variable Star Observers (AAVSO) database\footnote{https://www.aavso.org/} and $V$ observations from The All Sky Automated Survey (ASAS, \citealt{1997AcA....47..467P}). The typical error of observations in the ASAS and AAVSO data are 0.03~mag and 0.05~mag, respectively. The light curve is presented in Fig.~\ref{TCrB_all}. We searched for periodicities in the data using the discrete Fourier transform method of the $Period04$ program \citep{2005CoAst.146...53L}. For estimating the errors we used the \citet{1999DSSN...13...28M} method. The analysis of the AAVSO $B$ data gave the highest signal at the period of 114d, i.e. half of the orbital period, associated with ellipsoidal variability of the system. Besides that the highest signal was at 1077$\pm$19d. In the case of the AAVSO $V$ data the $\sim$1077d variability seems to be only marginally detected. In $VRI$ observations there is an additional signal present at P=87d and P=164d, both of which are one year aliases of the orbital period (Fig.~\ref{TCrB_powerspec}). We did not detect the $\sim$55d variability discovered by \citet{1988AJ.....95.1505L} that the authors attributed to pulsations of the red giant. The reality of the $\sim$55d variability was already discussed by \citet{1998MNRAS.296...77B} and the authors concluded that it was unlikely to be caused by red giant pulsations.

\subsubsection{Detection of flickering}
In order to detect flickering one must compare the measured scatter of the observations with expected uncertainty from CCD photometry. Usually this is done by using multiple observations to derive the parameters of the CCD camera (see e.g.  \citealt{2001MNRAS.326..553S}; \citealt{2006AcA....56...97G}). This method was not useful for our purposes since we did not have a sufficient number of observations collected with the same instrumental setup.  Another method is constructing a magnitude--standard deviation diagram for a large number of stars within the same exposure (see e.g. \citealt{1996AJ....111..414D}). This method was not fit for our observations since in our T~CrB field there were only two comparison stars. Therefore we derive our own method of estimating the expected uncertainty in observations of a star with a given magnitude, that requires only two comparison stars.

The error of an observed flux with a CCD camera is given by the equation
\begin{equation}\label{sigma1}
\sigma_i^2=\sigma_{CCD,i}^2+s^2\times N_i^2,
\end{equation}
where $i$ denotes the observed star, $\sigma_{CCD,i}$ is a standard uncertainty in CCD aperture photometry, $s$ is the ratio of rms flux variation to the mean flux due to turbulence in the atmosphere (\citealt{1963AJ.....68..395R};  \citealt{1967AJ.....72..747Y}) and  $N_i$ is the number of photons received from the star. The parameter $s$ is a function of air mass, height of the observatory, aperture of the telescope and exposure time. Therefore it should have approximately the same value for every star in the exposure. The uncertainty of the CCD photometry is given by the standard CCD equation (see e.g. \citealt{2000hccd.book.....H})
\begin{equation}\label{ccdeq1}
\sigma_{CCD,i}^2= N_i+n_{bins}\left( 1+\frac{n_{bins}}{n_{sky}}\right)(N_S+N_R^2+N_D),
\end{equation}
where $n_{bins}$ is the number of pixels used to estimate the flux of the star, $n_{sky}$ is the number of pixels used to estimate the level of background,  $N_S$ is the number of photons coming from the background, $N_R^2$ is the number of electrons from the readout noise and $N_D$ is the number of electrons from the  dark current. We carried out our photometry in such a way that $n_{bins}$ and $n_{sky}$ was the same for every star. Moreover, we assume that $N_R^2$ and $N_D$ are equal for every star observed, which is a good approximation for a modern CCD camera. Similarly, we assume that $N_S$ is the same for every star observed. We stress that this approximation is valid only for small field of view with negligible vignetting. We can rewrite equation (\ref{ccdeq1}) in a simpler form
\begin{equation}
\sigma_{CCD,i}^2= N_i+\xi.
\end{equation}
Equation \ref{sigma1} can now be rewritten as
\begin{equation}\label{ccdeq2}
\sigma_i^2=N_i+\xi+s^2\times N_i^2
\end{equation}
and rearranged to calculate $\xi$:
\begin{equation}\label{xieq}
\xi=\sigma_i^2-N_i-s^2\times N_i^2
\end{equation}
In order to estimate $\xi$ and $s^2$  we will use two standard stars, therefore we will change the $i$ index to 1 and 2 accordingly. We rewrite the equation~(\ref{xieq}) for the first and second standard star and compare the right sides of the equations
\begin{equation}
\sigma_1^2-N_1-s^2\times N_1^2=\sigma_2^2-N_2-s^2\times N_2^2
\end{equation}
After some algebra this gives
\begin{equation}\label{s2eq}
s^2=\frac{(\sigma_2^2-\sigma_1^1)-(N_2-N_1)}{N_2^2-N_1^2}
\end{equation}
combining equations (\ref{s2eq}), (\ref{xieq}) and (\ref{ccdeq2}), and  substituting $i$ in the  equation (\ref{ccdeq2}) to $v$, for the variable star, we get
\begin{equation}\label{verr}
\sigma_v^2=N_v-N_1+\sigma_1^2 + (N_v^2  -N_1^2 )  \frac{(\sigma_2^2-\sigma_1^2)-(N_2-N_1)}{N_2^2-N_1^2}
\end{equation}
This equation can be used to compare expected uncertainty of flux to the measured one in order to detect flickering.

\begin{table}
 \centering
  \caption{Summary of the flickering observations. The $\sigma_{\mathrm{measured}}/\sigma_{\mathrm{expected}}$ ratio is the ratio of measured error of mean photon count to the one calculated from Eq.~(\ref{verr}).  }\label{tableflickering}
  \begin{tabular}{|ccc|}
  \hline
Date & 03.08.2013 & 04.08.2013\\
MJD--mid & 56507.813  & 56508.839 \\
UT start--end & 19:07:36--21:31:30 &  18:41:10--20:20:30\\
mean U [mag] & 12.01 & 11.90 \\
U max--min [mag] & 11.839--12.132 & 11.729--12.036 \\
$\sigma_{\mathrm{measured}}/\sigma_{\mathrm{expected}}$ & 1.2 & 2.4\\
\hline
\end{tabular}
\end{table}

\subsection{X--ray observations}
Observations of T~CrB in the X--ray domain were carried out using the Rossi X--ray Timing Explorer (\citealt{1993A&AS...97..355B}; RXTE) Proportional Counter Array (\citealt{1996SPIE.2808...59J}; PCA). The PCA consists of five proportional counter units (PCUs). PCU0 and PCU1 are known to have lower quality observations at the end of the RXTE mission, therefore we used data from PCU 2, 3 and 4. Moreover we analyse data only from layer 1 of the PCUs, because the accuracy of background modelling for layers 2 and 3 is much lower. The observations were downloaded from the High-Energy Astrophysics Virtually ENlightened Sky database\footnote{www.isdc.unige.ch/heavens} \citep[HEAVENS;][]{2010int..workE.162W}. We extracted a light curve with a bin size equal to the exposure time of an individual observation for studying the long-term variability. Also extracted was a light curve with a bin size of 120s that we used for searching for short time variability. The bin size was selected in such a way that the lowest signal to noise ratio in 2-5~keV range at an individual poiting was hihger than 4.

\begin{figure}\centering
\resizebox{\hsize}{!}{\includegraphics{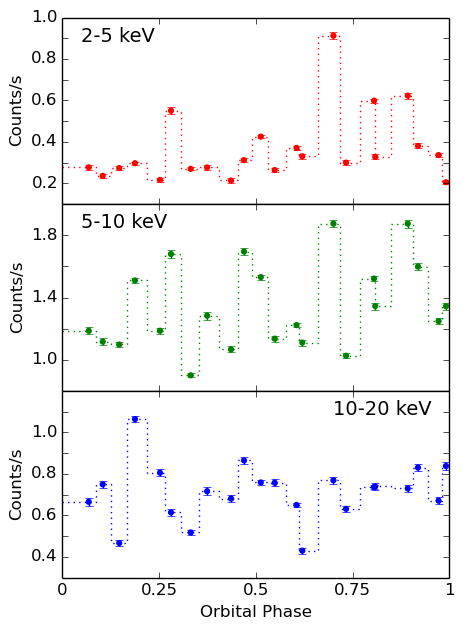}}
  \caption{Phase plot of RXTE PCA observations. The orbital phase was calculated using the ephemeris from  \citet{1988AJ.....95.1505L}. }
\label{TCrB_rxte_phase}
\end{figure}

\begin{figure}\centering
\includegraphics[width=.5\textwidth]{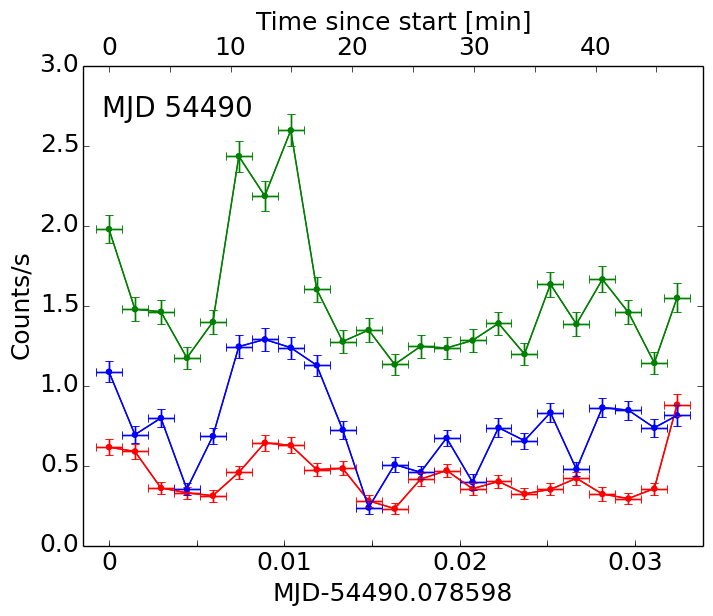}\\
\includegraphics[width=.5\textwidth]{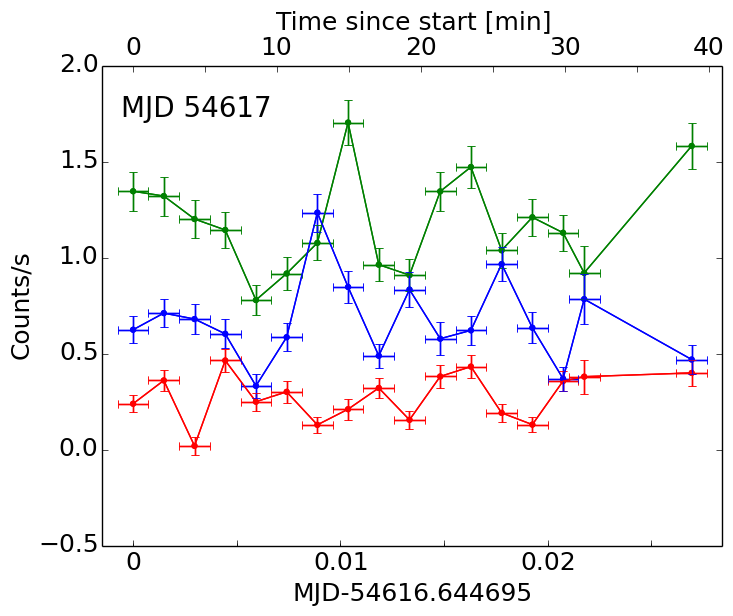}
  \caption{Selected observations of flickering in T~CrB in X--rays. The colours are the same as in Fig.~\ref{TCrB_all}.}
\label{rxte_flickering_individual}
\end{figure}

A log of observations is presented in Table~\ref{xraylog}. Additionally, we used $Swift$ observations that were published by \citet{2009ApJ...701.1992K}. The long term variability of T~CrB in X--rays compared to the optical observations is presented in  Fig.~\ref{TCrB_all}. A phase plot of RXTE data is presented in Fig~\ref{TCrB_rxte_phase}. The light curve with 120s bins from each individual pointing is presented in Figs.~\ref{rxte_flickering_individual}~and~\ref{rxte_flickering_individual2}.

\begin{table}
 \centering
  \caption{Log of RXTE PCA observations.}\label{xraylog}
  \begin{tabular}{|cccc|}
  \hline
Observation ID	&	Date	&	MJD	&	Exposure [ks]	\\
\hline
93007-03-01-00	&	02.07.2007	&	54283	&	3.045	\\
93007-03-02-00	&	16.08.2007	&	54329	&	2.980	\\
93007-03-03-00	&	23.09.2007	&	54367	&	3.345	\\
93007-03-04-00	&	02.11.2007	&	54407	&	3.105	\\
93007-03-05-00	&	14.12.2007	&	54449	&	2.965	\\
93007-03-06-00	&	25.01.2008	&	54490	&	2.926	\\
93007-03-07-00	&	07.03.2008	&	54533	&	3.212	\\
93007-03-08-00	&	20.04.2008	&	54576	&	3.010	\\
93007-03-09-00	&	30.05.2008	&	54617	&	3.059	\\
93007-03-10-00	&	11.07.2008	&	54658	&	3.694	\\
93007-03-11-00	&	22.08.2008	&	54700	&	2.914	\\
93007-03-12-00	&	03.10.2008	&	54742	&	2.975	\\
93007-03-13-00	&	14.11.2008	&	54785	&	2.805	\\
94007-03-01-00	&	26.12.2008	&	54826	&	2.970	\\
94007-03-02-00	&	09.02.2009	&	54871	&	3.073	\\
94007-03-03-00	&	23.03.2009	&	54914	&	2.915	\\
94007-03-04-00	&	02.05.2009	&	54954	&	2.859	\\
94007-03-05-00	&	13.06.2009	&	54995	&	3.414	\\
94007-03-06-00	&	25.07.2009	&	55038	&	3.338	\\
94007-03-07-00	&	06.09.2009	&	55080	&	2.819	\\
94007-03-08-00	&	16.10.2009	&	55120	&	2.820	\\
94007-03-09-00	&	28.11.2009	&	55163	&	3.602	\\
\hline
\end{tabular}
\end{table}

We attempted spectral analysis of the data using XSPEC version 12.9.0. Following the work of \citet{2009ApJ...701.1992K} and \citet{2014MNRAS.437..857E} a single-temperature bremsstrahlung emission together with  a Gaussian line to account for the Fe~K$\alpha$ fluorescence line subject to full-covering and partial-covering absorption were employed. The obtained fits were very good (reduced $\chi^2\simeq0.95$), but physical parameters from our fit had errors too large for an in-depth  analysis. Nonetheless, in order to investigate  the spectral evolution of the star we show spectra from two epochs, i.e., the first epoch was set before MJD~54650, in which T~CrB showed flares in the soft band (see Fig.~\ref{TCrB_all}) of the spectrum; the second epoch  was set after MJD 54650 where a flare in the hard band (see Fig.~\ref{TCrB_all}) of the spectrum occurred, while the soft component remained quiet.

\begin{figure}\centering
\resizebox{\hsize}{!}{\includegraphics{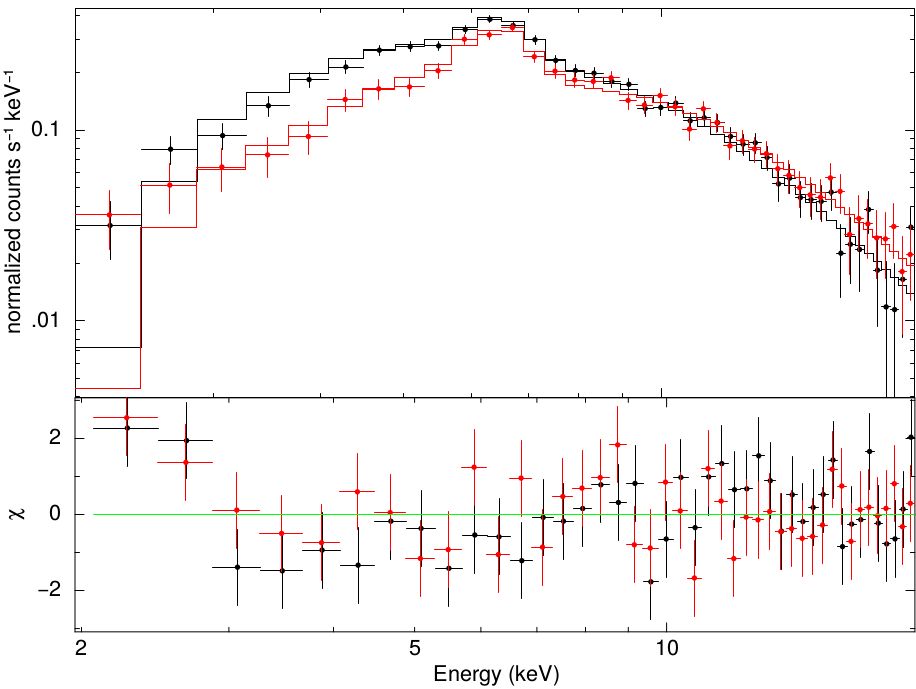}}
\caption{Top panel: Folded model with an absorbed bremsstrahlung plus a Gaussian (solid lines) and observed spectra (points) for MJD<54650 (black) and MJD>54650 (red). Bottom panel: the corresponding residuals.}
\label{rxte_fitplot}
\end{figure}

The results of the spectral fitting are listed in Table~\ref{xrayfitparams} and shown in Fig.~\ref{rxte_fitplot}, indicating a possible tail of a soft excess in the spectrum at energies below 3~keV. This excess was not detected in previous observations of T CrB (\citealt{2008ASPC..401..342L}; \citealt{2009ApJ...701.1992K}; \citealt{2013A&A...559A...6L}; \citealt{2014MNRAS.437..857E}). If confirmed, it would change the classification of this system to $\beta/\delta$ type according to \citet{1997A&A...319..201M} and extended by \citet{2013A&A...559A...6L}.


\begin{table*}
 \centering
  \caption{Spectral analysis from the RXTE/PCA spectra of T CrB using a single-temperature, absorbed bremsstrahlung spectral model with a Gaussian line to account for the Fe~K$\alpha$ fluorescence line.}\label{xrayfitparams}
  \begin{tabular}{|cccccccccc|}
  \hline
Epoch	&	$N_H$(FC)$^{\rm a}$	&	$N_H$(PC)$^{\rm b}$	&	PCF$^{\rm c}$ & kT  & Line Peak & Line Width 	& $\chi^2$/(d.o.f.) \\
   & [$\times10^{22}$cm$^2$]& [$\times10^{22}$cm$^2$] & & [keV] & [keV]& [keV] \\
\hline
MJD<54650	&	<1.7	&	30.5$^{+12.7}_{-11.4}$	&	0.80$_{-0.08}^{+0.06}$ & 31$_{-15}^{+84}$ & 6.53$_{-0.14}^{+0.15}$ & 0.28$_{-0.28}^{+0.35}$ & 0.98\\
MJD>54650	&	<3.2	&	41.7$^{+17.1}_{-14.4}$	&	 0.88$_{-0.05}^{+0.05}$ & 52$_{-34}^{+52}$ &  6.53$_{-0.15}^{+0.15}$ & 0.21$_{-0.21}^{+0.38}$ & 0.93\\
\hline
\end{tabular}

\raggedright
\textbf{Notes}: $^{\rm a}$~Column density of the material covering the whole source; $^{\rm b}$~Column density of the material partially covering the source; $^{\rm c}$~Partial covering fraction.
\end{table*}

\section{Results}

\subsection{Active phases}\label{active_phase_section}

\subsubsection{Photometric variability}\label{photvar_sec}

The active phases of the hot component in T~CrB have been studied for many years (see e.g. \citealt{2004A&A...415..609S} and references therein). However, the recurrent time-scale of the active phases is not clear. On the basis of 40 years of visual observations, \citet{1997MNRAS.287..634L} estimated that the active phases occur with a period of $\sim$9840d. However, the observations analysed by \citet{1997MNRAS.287..634L} were done by eye and therefore may have been more sensitive to variability of the RG, which dominates the visual spectral range.  Moreover, the $\sim$9840d period was not confirmed by more recent observations, most likely due to the observations covering an insufficient time baseline. Thus, the $\sim$9840d period remains uncertain. 

Observations collected with the International Ultraviolet Explorer (IUE) during 1978--1990 revealed two minima in the UV continuum separated by $\sim$5000d \citep{1992ApJ...393..289S}. Similarly, the $U$ light curve in 1982--2001 shows two maxima with an amplitude of $\sim$2~mag separated by $\sim$5000d \citep{2004A&A...415..609S}. In addition, variability in the emission line fluxes with a similar period was reported by \citet{1997ppsb.conf..117A}. Therefore, we conclude that the most prominent active phases in T~CrB occur with a period of $\sim$5000d (hereafter `big active phases').

Additional activity of the hot component was detected with a period of $\sim$1000d by \citet{1997IBVS.4461....1Z}, where the authors observed three brightenings in the $U$ filter with an amplitude of $\sim$1~mag separated by $\sim$1000d at MJD 48100, 49100 and 50200. The variability reported by \citet{1997IBVS.4461....1Z} was however during a minimum of the big active phase \citep[see e.g. fig.~4 in][]{2004A&A...415..609S}. Variability of a similar time-scale was detected in the emission lines \citep{1997ppsb.conf..117A}. Therefore active phases with a smaller amplitude and a time-scale of $\sim$1000d are present as well (hereafter `small active phases'). The small active phases have had been observed only during a minimum of the big active phases (e.g. in the UV continuum in observations presented by \citealt{1992ApJ...393..289S} or in $U$ photometric observations presented by \citealt{2004A&A...415..609S}). This suggest that both the big and small active phases are connected.

The variability with a period of $\sim$1000d is also present in our observations (Fig. \ref{TCrB_all}; Fig~\ref{TCrB_powerspec}). This variability is particularly prominent in $B$ as expected for activity associated with the hot component. The amplitude and the time-scale are consistent with the small active phases. 

During most of the period of our observations the magnitude of T~CrB, $B \ga 11$, i.e. the system has remained in a minimum of the big active phase \citep[see e.g. fig.~4 in][]{2004A&A...415..609S}, and only after MJD~57100 did the system begin to brighten. This suggests that our observations in the $B$ filter mainly covered a period of small active phases and ended at the beginning of a rise to the maximum of a big active phase. This rise is corroborated by archival Swift UV photometry that also show T~CrB brightening during the same period (see Appendix \ref{sec:uvot}).

\subsubsection{Emission line variability}\label{specsection}

The correlation between H$_\alpha$  fluxes and the active phases was discussed by \citet{1990JAVSO..19...28I}, \citet{1991MNRAS.253..605A}, \citet{1999A&A...344..177A} and \citet{2004A&A...415..609S}. The authors reached a conclusion that the line strengths are correlated with the brightness of the hot component. Most notable was the variability of emission lines with a period of 3640d and 906d \citep{1997ppsb.conf..117A}.

In the past, the most thorough studies of emission line variability concentrated on their equivalent widths (EW) \citep[e.g.][]{1997ppsb.conf..117A}, while the studies of emission line fluxes were much more sporadic \citep[e.g.][]{1999A&A...344..177A}. The downside of using EW is that the continuum can be influenced by hot component variability, especially at short wavelengths (e.g. around \mbox{He\,{\sc ii}\,4686}). In this study, we pay more attention to the variability of emission line fluxes. 

From Fig.~\ref{all_fluxes} and Fig~\ref{TCrB_all} it is clear that the fluxes of all of the measured emission lines are correlated with the photometric observations, which is consistent with the previous studies. The EW of H$_\alpha$ during the first part of our observations was $<15\AA$, which points to the minimum of the big active phase \citep[see e.g.][]{2004A&A...415..609S}. The increase of the emission line fluxes and their EW at the last observations suggest we are approaching the maximum of a big active phase, which was already evident in photometric observations.

The ratio of F(H$_\alpha$)/F(H$_\beta$) changes from $\sim 3$ to 13. The reddening free ratio F(H$_\alpha$)/F(H$_\beta$) $\sim 3-10$ is typical for SySt, and it is due to self-absorption effects \citep{1996ApJ...471..930P,1997A&A...327..191M}. In particular, F(H$_\alpha$)/F(H$_\beta$) showed a short maximum around MJD 56750. Simultaneously there was a minimum in both photometric observations and line fluxes. Moreover, the observed increase of the F(Ha)/F(Hb) ratio was during a minimum of the small active phase as well.  These changes are most likely due to variable self-absorption and optical depth effects as indicated by the Ha and Hb profiles being heavily affected by an absorption component seen in most SySt (see e.g. \citealt{1993A&AS..102..401V}).

\subsubsection{X--ray variability}\label{xraysection}

\begin{figure}\centering
\resizebox{\hsize}{!}{\includegraphics{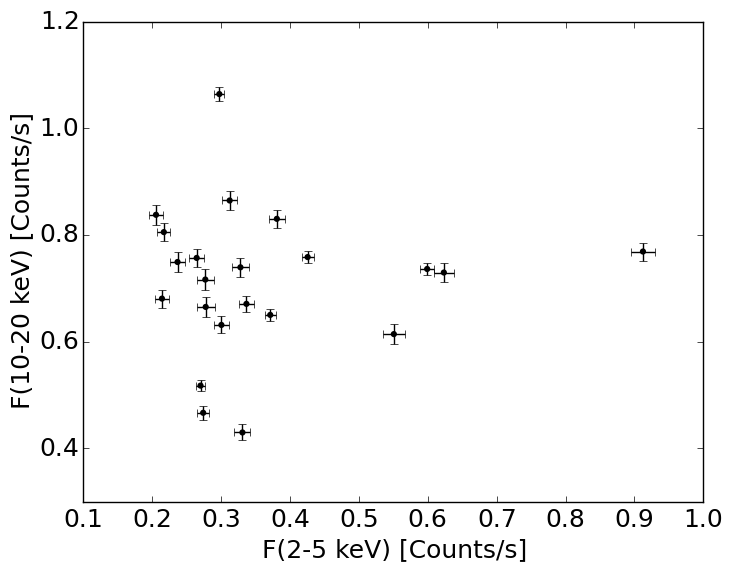}}
  \caption{Relation between the count rate in the hard and soft parts of the X--ray spectrum in RXTE observations.}
\label{TCrB_hardness}
\end{figure}

In the case of X--ray emission the variability in the soft range (2--5 keV) is not correlated with variability in the hard range (10--20 keV; Fig.~\ref{TCrB_hardness}). Furthermore, the variability in the intermediate range (5--10 keV) seems to be a superposition of the variability in the hard and soft range. This indicates two separate sources of the X--ray variability. In line with this interpretation is the variability in both hard and soft X--rays of T CrB reported by \citet{2009ApJ...701.1992K}. However, their observations in both spectral ranges were not contemporaneous, so they cannot directly confirm our interpretation. 

The variability in the soft range is possibly related to changes in the column density of absorbing material or variable partial-covering fraction, as was observed in RT~Cru and suggested for T~CrB by \citet{2009ApJ...701.1992K}. This kind of variability would be observed in the soft and intermediate ranges, but would not affect the hard part of the spectrum, which is consistent with the X--ray light curve. On the other hand, variability in the hard range would reflect variability of the source of X--ray radiation. The X--ray photons in T~CrB are thought to originate in the boundary layer between the accretion disc and the WD \citep{2008ASPC..401..342L}, so this band may be tracing variable accretion.

It is clear that the X--ray variability is unrelated to the orbital phase (Fig~\ref{TCrB_rxte_phase}) and to UV changes. Most probably the X--ray variability is related to the active phases discovered in the visible bands.  Due to the short time coverage of the RXTE observations it is impossible to study the periodicity of the changes in X--rays over the duration of the big and small active phases.

In the RXTE observations there are two maxima in the soft band at MJD~54329 and MJD~54533, followed by a period of relative quiescence (Fig.~\ref{TCrB_all}). The hard part of the spectrum showed a shallow minimum at MJD~54407 between these two maxima of the soft component. Moreover, in the hard band around MJD~54742 there was a minimum followed by a maximum around MJD 54871. In the time during the maximum in the hard band there were no significant changes in the soft band. Generally it seems that the maxima in the soft band are during a  quiescence in the hard band and vice versa, but the number of observed maxima is too low for a meaningful conclusion to be drawn. If this hypothesis holds, then this could mean that the variability in both bands is a different phase of the same phenomenon, such as e.g. disc instability or increased mass loss from the RG.

In order to investigate the possible periodicity in the X--ray variability we utilised Swift observations from \citet{2009ApJ...701.1992K}. These observations are in the band 14--24~keV and therefore the changes in the count rate are associated with the hard band of RXTE. The count rate during the Swift observations showed similar behaviour as in the RXTE observations around MJD 54800. Namely it showed a period of relative quiescence with $7\times10^{-5}$ counts/s/detector followed by a short minimum with $5\times10^{-5}$ counts/s/detector succeeded by an immediate rise to the maximum reaching $12\times10^{-5}$ counts/s/detector.

The similarity of behaviour in the Swift and RXTE observations could point to a periodic or quasi-periodic nature of the variability. The two maxima in the Swift and RXTE observations are separated by $\sim$1200d. Moreover there is only one maximum in the hard band during the period of 880d covered by RXTE. This hints that the observed variability is associated with small active phases discussed in section~\ref{photvar_sec}. On the other hand, two maxima in the soft band took place around the time of a minimum of small active phases. Therefore the soft and hard X--ray variability could have the same period, but with a phase shift. The whole period covered by Swift and RXTE was during a quiescent state of the big active phases.

\subsection{Flickering}\label{flickering_section}

\subsubsection{Optical observations}

\citet{2004MNRAS.350.1477Z} showed that the amplitude of flickering in T~CrB is correlated with the flux of the blue part of the spectrum. A similar relation was discovered in other symbiotic stars -- CH~Cyg \citep{1990AcA....40..129M}, MWC~560 \citep{1996A&AS..116....1T} and RS~Oph \citep{2015MNRAS.450.3958Z}. Most of the observations used by \citet{2004MNRAS.350.1477Z} were obtained during a period of maxima of two big active phases and only two from the quiescent epoch  (section~\ref{photvar_sec}) and only two from the time of quiescence (see their fig.~2). Therefore, it is clear that the flickering amplitude changes are connected with big active phases, but it is not clear how a small active phase can influence the flickering. 

Interestingly, the observations reported by \citet{2010ATel.2586....1Z} from MJD 54851 and by \citet{2015AN....336..189Z} from MJD 55316 are from the time of a minimum of the big active phase variability, but the amplitude of the flickering was $\sim0.4$~mag in $U$, which is larger than the mean amplitude of the flickering reported by \citet{2004MNRAS.350.1477Z} observations ($\sim$0.35~mag in $U$), but significantly lower than their maximum amplitude (0.6~mag in $U$). At these times the system was close to the maximum of a small active phase. Moreover, \citet{2015AN....336..189Z} reported a flickering with a very low amplitude  of 0.17~mag in $U$ at MJD 55603, which was during the minimum of a small active phase. This amplitude is lower than any of the reported by \citet{2004MNRAS.350.1477Z}, where the lowest amplitude was 0.22~mag.  Similarly, the amplitudes of the flickering of $\sim$0.3~mag we determined in section~\ref{sec:phot} are also from the time of another quiescent state of a big active phase and is close to the lower values reported by \citet{2004MNRAS.350.1477Z}. While the scarcity of the flickering observations during the minimum of big active phases is making the conclusions uncertain, the observations suggest that the small active phases follow the same flux--amplitude of the flickering relation as during the big active phases. This points to the same physical process behind the big and small active phases.

\subsubsection{X--ray observations}

X--ray flickering in T~CrB was discovered by \citet{2008ASPC..401..342L}, who reported variability on time scales of minutes. We discovered similar flickering in most of the RXTE observations. From the individual lightcurves it is clear that the amplitude of the flickering is larger in the hard band than in the soft band, where it is marginally detected (Fig.~\ref{rxte_flickering_individual},\ref{rxte_flickering_individual2}). This is consistent with the hypothesis that the hard band traces variability of accretion through the boundary layer, where the flickering is expected to originate.  

\citet{2015A&A...579A..50B} discussed a time lag in the observations of flickering at different wavelengths in CVs on the basis of optical observations. The author discovered that in most systems the flares reach their maxima slightly earlier at shorter wavelengths. Our observations had too short a timebase for a meaningful cross--correlation analysis. At some epochs the flickering in the 5--10~keV range seems to be correlated with the flickering in 10--20~keV range within our time resolution. The best examples are observations at MJD 54490 and MJD 55163 with correlation coefficients between the 10--20~keV and 5--10~keV flux of r=0.83 and r=0.76, respectively. On the other hand, on MJD 54616 there was a flare in the 10--20~keV range which reached a maximum $\sim$12~min after the start of the observation and the corresponding flare in the 5--10~keV range reached its maximum $\sim$2~min later. On MJD 54407, a similar lag of $\sim$2~min was observed in a flare that reached maximum in the 10--20~keV range $\sim$42~min after the start of observation. 

At other times the variations in the hard range are not correlated at all with the flickering in 5--10~keV range (see e.g. observations at MJD 54826). Most interestingly at some times the variations in the 5--10~keV range and the 10--20~keV range are anti--correlated. This can be seen e.g. between 30 and 35~min of observations at MJD 54913 (r=0.25) or in two brightenings at MJD 54616 (r=0.02) between 20 and 40~min. This suggests a change in the hardness ratio of the flickering rather than the total amplitude.  

Overall the flickering seems to be most prominent in the hard band, but the hardness ratio of flares can change during the observing run. Some flares can have very hard spectra and are only observed in the 10--20~keV range (e.g. at MJD 54826 with r=0.27) and in some rare cases they can have softer spectra and are observed in the 2--10~keV range and not in the 10--20~keV range (see e.g. a flare in the first 5~min of observations at MJD 54995). This seems to be consistent with the fact that in the optical observations the source of the flickering has a different colour at different times \citep[see e.g. tab~2 in][]{2015AN....336..189Z}. Nevertheless, the source of the flickering in optical observations and in the X--rays could not be the same since in the optical the source typically has a temperature of $\sim$9000~K in symbiotic recurrent novae \citep{2015AN....336..189Z}, which seems consistent with reprocessing in the surrounding \mbox{H\,{\sc ii}} region.

\begin{figure}\centering
\resizebox{\hsize}{!}{\includegraphics{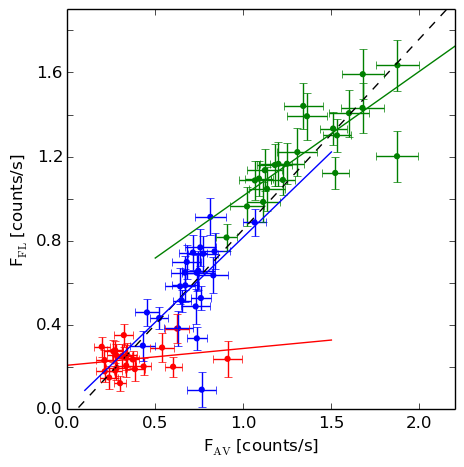}}
  \caption{The relation between brightness and flickering amplitude in the X--ray observations. The error is the mean error in count rate measure during one pointing of the telescope. The solid line is a linear fit to the data from a spectral range represented by the same colour. Black dashed is a linear fit to data points from all spectral ranges simultaneously. The colours are the same as in Fig.~\ref{TCrB_all}.}
\label{flickering_amplitude}
\end{figure}

We studied the total flux--flickering amplitude relation in the X--ray observations. For this we plotted an average count rate (F$_{\mathrm{AV}}$) versus the average amplitude of the flickering (F$_{\mathrm{FL}}$=F$_{\mathrm{max}}$--F$_{\mathrm{min}}$). We calculated the Pearson's correlation coefficient $r$ and probability $p$ for a hypothesis test whose null hypothesis is that the slope of the relation is zero. The calculated values are listed in Tab.~\ref{X--ray_flickering_tab} and the data together with the fit are shown in Fig.~\ref{flickering_amplitude}. From the analysis it is clear that the correlation is present in the 5--10~keV and 10--20 keV range and absent or marginally present in the 2--5~keV range. The correlation is stronger in the 5--10 keV range rather than in the 10--20 keV range. We attribute this to the fact that the fit was carried out to data spanning over a larger range of F$_{\mathrm{AV}}$ in the 5--10 keV range rather than in the 10--20 keV range. The best fit is obtained using data points from all of the spectral ranges simultaneously. This suggests that the  total flux--flickering amplitude relation is the same over the whole spectral range. Since the count rate is proportional to the physical flux we can compare the slope of the fit to a similar fit in the optical range. The slope in the $U$ band for the F$_{\mathrm{AV}}$--F$_{\mathrm{FL}}$ relation is 0.156$\pm$0.042 for T~CrB \citet{2004MNRAS.350.1477Z}. Since the values for optical and X--ray observations are not consistent this is another indication that the flickering in those two wavelength bands originates from a different place. However, we note that if the relation is logarithmic rather than linear, then the optical and X--ray observations may follow the same relation \citep{2016MNRAS.457L..10Z}

There are only two observations of flickering in the optical range during the period covered by X--ray observations and unfortunately none of them were simultaneous with the X--ray observations. \citet{2010ATel.2586....1Z} observed flickering at MJD 54851 and MJD 54889 with amplitudes in $U$ 0.404~mag and 0.303~mag respectively, which are rather high for flickering outside of the big active phase in T~CrB. These observations were taken close to the maximum of the hard X--ray emission, which seems to confirm that the hard X--ray and optical flickering are connected.

\begin{table}
\caption{Quantities of the linear fit to the F$_{\mathrm{av}}$--F$_{\mathrm{fl}}$ relation in the X--ray observations. The combined fit is a simultaneous fit to data points from all of the spectral ranges.}
\label{X--ray_flickering_tab}
\begin{tabular}{|ccccc|}
\hline
Range 	&	2--5 kev & 5--10 keV & 10--20 keV & Combined	\\
slope	&	0.08$\pm$0.08 & 0.6$\pm$0.1 & 0.8$\pm$0.3 & 0.90$\pm$ 0.04\\
$r$ 	& 0.22 & 0.80 & 0.55 & 0.93 \\
$p$     &  0.32 & 8.0$\times$10$^{-6}$ & 7.9$\times$10$^{-3}$ & 3.7$\times$10$^{-30}$\\
\hline
\end{tabular}
\end{table}

\section{Discussion}\label{discussion_sec}

\subsection{Active phases}

Similar active phases to those observed in T~CrB have been discovered in the other SyRN RS~Oph \citep{2008ASPC..401..219G}. Namely, the authors observed brightenings of the system on time-scales of 1200--1800d with an amplitude of $\sim$1~mag. \citet{2008ASPC..401..219G} pointed out that the behaviour of RS~Oph outside of the nova outbursts resembles the classical Z~And variability, which is thought to be a result of an instability in the accretion disc (\citealt{2003ASPC..303....9M}; \citealt{2006ApJ...636.1002S}). The active phases in T~CrB and RS~Oph could have the same origin as in the Z~And type symbiotic systems with the difference that in the Z~And type systems there is steady hydrogen burning on the surface of the WD. 

If the active phases in T~CrB and RS~Oph are driven by the same mechanism as the multiple-outburst activity of Z~And type SySt, similar time-scales should be expected. In fact, outbursts in many Z~And type systems recur with quasi-periods of one to tens of years. 

Namely, the major outburst in AG~Dra occurs every 12--15 years, while minor outbursts are observed roughly once a year \citep{2016MNRAS.456.2558L}. Z~And showed a time-scale of $\sim$8400d of the outburst activity \citep{1994A&A...292..534F} and the interval between outbursts in BF~Cyg is $\sim$6376d \citep{2006MNRAS.366..675L}. Given the differences between the systems these time-scales are in rough agreement with the ones in T~CrB and RS~Oph. 

A useful comparison can also be made based on similar physical parameters of these systems. In the prototype star Z~And, after assuming M$_{\mathrm{WD}}\sim$0.65M$_\odot$ \citep{1997A&A...327..219S}, a hot component mass function of f(m)=0.0240M$_\odot$ \citep{2000AJ....120.3255F}, an inclination of i=$41^\circ$ \citep{2010AJ....140..235I}, and orbital period of 759d \citep{2000AJ....120.3255F}, the radius of the Roche Lobe around the WD is $\sim$140R$_\odot$. Similarly, for T~CrB, assuming M$_{\mathrm{WD}}=1.2$M$_\odot$ and M$_{\mathrm{RG}}=0.8$M$_\odot$ \citep{1998MNRAS.296...77B,2004A&A...415..609S}, and an orbital period of 227.67d \citep{1988AJ.....95.1505L}, the radius of the Roche Lobe around the WD is $\sim$80R$_\odot$. In the case of RS~Oph, taking M$_{\mathrm{WD}}=1.3$M$_\odot$, M$_{\mathrm{RG}}=0.74$M$_\odot$, and an orbital period of 453.6d \citep{2009A&A...497..815B}, the radius of the Roche Lobe around the WD is $\sim$135R$_\odot$. The estimated mass transfer rate in T~CrB is 2.5$\times 10^{-8}$~M$_\odot$/yr \citep{1992ApJ...393..289S}, whereas in Z~And the mass transfer rate varies between 4.5$\times 10^{-9}$~M$_\odot$/yr during quiescence \citep{1988ApJ...324.1016F} to 3.2$\times 10^{-7}$~M$_\odot$/yr during the active phase \citep{2004A&A...428..985T}. Therefore, the mean mass transfer rates are of the same order of magnitude in both systems.

The ratio of Roche Lobe radii around the WDs in Z~And and T~CrB is $\sim$1.5. It is interesting to note that this value is similar to the ratio of $\sim$1.7 between time-scales of Z~And outburst activity and big active phases in T~CrB. On the other hand, the ratio of Roche Lobe radii around the WDs in RS~Oph and T~CrB is $\sim$1.7, which is similar to the ratio of $\sim$1.5 between time-scales of active phases in RS~Oph and small active phases in T~CrB. This is to be expected, because the time-scale of a disc instability scales with the size of the accretion disc, which in turn is proportional to the radius of the Roche Lobe around the WD. This is more consistent with the longer time-scale of outburst activity in Z~And than the time-scale of active phases in T~CrB.

Recently,  disc instability in RS~Oph  has been modeled by \citet{2011MNRAS.418.2576A}, who found recurring outbursts in the system on 10--20 yr time-scales, which is comparable to the time-scales of  active phases actually observed in RS~Oph and Z~And \citep{2008ASPC..401..219G}. In some models alternating small and big outbursts take place. This could be associated with big and small active phases in T~CrB.


\citet{1997ppsb.conf..117A} proposed a solar--type cycle resulting in a variable mass transfer as an explanation of the active phases in T~CrB. In order to obtain solar-like cycles one must have a high magnetic field. In the case of T~CrB this is probable since it is argued that in SySt the RG is more magnetically active than a single RG \citep{2002MNRAS.337.1038S}. In general, the hypothesis that the RG is a source of the observed activity is particularly interesting since variability with a time-scale of $\sim$1000d was observed in a single semi--regular variable WZ~Cas \citep{2005A&A...440..295L}.

Overall it seems to be clear that at least the big active phases are related to a change in mass transfer rate. Therefore it is important to determine the exact nature of the T~CrB variability, especially since SySt are thought to be promising SNIa progenitors \citep{2013IAUS..281..162M}.

Recently, \citet{2016NewA...47....7M} have claimed that during 2015 T~CrB  displayed ''super-active conditions never seen before''. They have also suggested that this super-active state is a new form of activity and it is different from the active phases observed e.g. by \citet{1990JAVSO..19...28I}. We argue that this is not true, and the activity observed by \citet{2016NewA...47....7M} is just a new maximum of the big active phases (see Sec.~\ref{active_phase_section}) and similar activity has been observed in the past.

\citet{2016NewA...47....7M} based their claim on a large increase in emission line fluxes and optical magnitudes of the system. However, as the authors noted, it  is difficult to study the history of emission line variability, since the observations are sparse and the authors rarely have given line fluxes. Here we expand the analysis of \citet{2016NewA...47....7M} using EWs. From Fig.~\ref{all_fluxes} one can see that variability of emission line fluxes is well represented by variability of EW(H$\alpha$). In particular, the $H\alpha$ emission line flux, estimated on the spectrum from MJD~57447, is F($H\alpha$)=4.5$\times$10$^{-11}$~erg~cm$^{-2}$~s$^{-1}$ , and the corresponding corresponding EW($H\alpha$)=48.8. This is much higher than the maximum flux F($H\alpha$)=3.9$\times$10$^{-11}$~erg~cm$^{-2}$~s$^{-1}$ observed by \citet{2016NewA...47....7M}. Therefore we adopt EW($H\alpha$)=48.8 as a maximum EW observed during the supposed super-active conditions. 

During the past big active phases, the EW($H\alpha$) varied from $\sim$5 to $\sim$40 \citep[see  fig.~4 in][]{2004A&A...415..609S}. Moreover, the maximum EW observed during the 1989 big active phase, EW(H$\alpha$)$\sim$40, was bigger than that observed during the 1997 big active phase (EW(H$\alpha$)$\sim$30). This shows that the amplitude of emission line variability changes, which is consistent with the quasi-periodic nature of the known phenomenon rather than a new form of activity. This is further supported by the photometric variability observations reported by \citet{2016NewA...47....7M} with timescales and amplitudes consistent with previous big active phases \citep[see  fig.~4 in][]{2004A&A...415..609S}.

\subsection{Flickering}

Since the flickering in T~CrB is mainly observed in the hard X--rays, which are thought to originate in the accretion disc boundary layer, our study seems to confirm the model of flickering originating from an unstable mass transfer through the boundary layer. However, it seems that  more than one mechanism is responsible for the observed variability. The events in which there is brightening in the 5--10~keV range and fading in the 10--20~keV range seem to indicate Compton cooling. Moreover, it is obvious that the flickering observed in the optical range is not the same as that observed in the X--rays since the optical colours of the flickering source in T~CrB are consistent with a blackbody of $\sim$9000~K \citep{2015AN....336..189Z}. Nevertheless, it seems that flickering in the hard X--rays  is related to the optical flickering since the maximum in hard X--rays was observed close to the maximum of the optical $\sim$1000d active phase, and for both wavelengths the amplitude of the flickering--mean flux relation holds. Our results seem to be best reproduced by a model proposed by \citet{2014MNRAS.438.1233S}  in which the flickering observed in the visible range comes from X--ray reprocessing by a geometrically thick disc. This model provides a natural explanation for the difference between the flickering source colours in SySt and classical CVs \citep{2015AN....336..189Z} as the accretion disc in SySt is thought to be bigger than in CVs. Reprocessing of the X--ray radiation by the nebula is in principle not excluded, but reprocessing by a thick accretion disc seems likely since the UV to IR SED of the system was successfully modelled by \citet{2004A&A...415..609S} using a model containing an optically thick accretion disc. On the other hand, reprocessing of the radiation by a nebula would be consistent with flickering observed in emission lines of SySt \citep{2005PASP..117..268Z,2014MNRAS.442.2637W}.

\section{Summary}\label{sumarysec}

We analysed optical, UV and X--ray observations of active phases and flickering in T~CrB. The main results of our work include:

\begin{itemize}
\item We found that there are big and small active phases of the hot component. Big active phases occur with a period of $\sim$5000d, while small active phases occur with a period of $\sim$1000d during a quiescence of the big active phases.
\item Variability in the soft part and hard part of the X--ray spectrum of T~CrB is not correlated. Variability in the soft band is consistent with a variable amount of absorbing material in the line of sight. Variability of the hard part of the spectrum most probably traces variable accretion through the boundary layer. Variability in both spectral ranges is not correlated with orbital phase, but seems to be correlated with active phases seen in the optical. 
\item The flickering seems to be present only in the hard X--ray component. The flickering in this spectral range follows a similar mean flux--amplitude of the flickering relation as the one observed in the optical. 
\item The different behaviour of flickering in X--rays observed at different times seems to point at more than one physical process behind flickering. The most likely scenario suggested by our observations is formation of the flickering in the boundary layer, which is then reprocessed from X--rays to the visible light by the thick accretion disc or a nebula around the system.
\item Time-scale of active phases in T~CrB is similar to time intervals between outbursts in Z~And type SySt. This suggests that the mechanism driving the activity in these systems is the same, presumably unstable disc accretion.

\end{itemize}

To understand the nature of the system further it would be most beneficial to acquire simultaneous X--ray, UV and optical observations of the flickering behaviour. This would help connect the X--ray and optical flickering. Moreover, the long--term X--ray monitoring would help to understand the X--ray behaviour of the system during the active phases which would lead to understanding the nature of the mass transfer variability in the system. This, in turn, could help understand the variability of Z~And type systems.

\section*{Acknowledgements}
We are grateful to all of the amateur astronomers that contributed their observations to this paper. In particular, we are thankful to members of the ARAS group for their wonderful work. We acknowledge with thanks the variable star observations from the AAVSO International Database contributed by observers worldwide and used in this research. We would like to thank the anonymous referee for the constructive comments that improved the quality of our manuscript significantly.

KI has been financed by the Polish Ministry of Science and Higher Education Diamond Grant Programme via grant 0136/DIA/2014/43. This study has been partially financed by Polish National Science Centre grant 2015/18/A/ST9/00746.  KS gratefully acknowledge observing grant support from the Institute of Astronomy and Rozhen National Astronomical Observatory, Bulgarian Academy of Sciences.




\bibliographystyle{mnras}
\bibliography{literature} 



\appendix
\label{sect:appendix} 

\section{Additional material}

\begin{landscape}
\begin{table}
 \centering
  \caption{ Log of spectroscopic observations from the ARAS database. }\label{logspec}
  \begin{tabular}{|ccccccccc|}
  \hline
Date	&	UT	&	MJD	&	Observer	&	Site	&	Instrument	&	Resolution	&	Range [\AA]	&	Exposure [s]	\\
\hline																	
2010-05-12	&		&	55328	&		&		&		&		&	4327 - 7290	&		\\
2010-08-02	&		&	55410	&		&		&		&		&	4510 - 7158	&		\\
2012-04-01	&	22:01	&	56018	&	S. Charbonnel	&	Durtal(UAI949)	&	T500F6-AP3600Eshel	&	10000	&	4277 - 7120	&	3626	\\
2013-02-17	&	3:01	&	56340	&	F. Teyssier	&	Rouen	&	SC25cm+LISA	&	967	&	4001 - 7000	&	3648	\\
2013-04-02	&	2:14	&	56384	&	F. Teyssier	&	Rouen	&	SC25cm+LISA	&	751	&	4000 - 7550	&	1210	\\
2013-04-03	&	0:32	&	56385	&	C. Buil	&	Castanet	&	CN212ALPY600ATIK4	&	617	&	3741 - 7561	&	3341	\\
2013-04-18	&	2:56	&	56400	&	F. Teyssier	&	Rouen	&	SC25cm+LISA	&	773	&	4000 - 7501	&	1210	\\
2013-05-02	&	23:19	&	56415	&	F. Teyssier	&	Rouen	&	SC25cm+LISA	&	757	&	3800 - 7500	&	1209	\\
2013-05-06	&	23:08	&	56419	&	F. Teyssier	&	Rouen	&	SC25cm+LISA	&	801	&	4001 - 7500	&	1637	\\
2013-06-04	&	20:47	&	56448	&	C. Buil	&	Castanet	&	C11ALPY600ATIK640	&	624	&	3750 - 7395	&	2463	\\
2013-07-19	&	20:55	&	56493	&	F. Teyssier	&	Rouen	&	SC25cm\_LISA\_ATIK46	&	836	&	4000 - 7501	&	1535	\\
2013-07-20	&	21:04	&	56494	&	F. Teyssier	&	Rouen	&	SC25cm\_LISA\_ATIK46	&	851	&	4001 - 7500	&	2421	\\
2013-08-01	&	20:19	&	56506	&	F. Teyssier	&	Rouen	&	SC25cm\_LISA\_ATIK46	&	817	&	4001 - 7500	&	1843	\\
2013-08-09	&	20:49	&	56514	&	F. Teyssier	&	Rouen	&	SC25cm\_LISA\_ATIK46	&	811	&	4001 - 7500	&	2151	\\
2014-03-20	&	2:03	&	56736	&	J. Montier	&	C.A.L.C	&	Meade355mm+Atik460E	&	601	&	3813 - 7391	&	1843	\\
2014-04-03	&	5:48	&	56750	&	T. Lester	&	Arnprior\_ON\_CA	&	311DK6.6+23um600+QS	&	1345	&	4024 - 7482	&	5099	\\
2014-04-05	&	22:44	&	56753	&	Ch. Revol	&	OHP	&	AP160-LISA	&	765	&	3901 - 7400	&	2413	\\
2014-04-06	&	00:21	&	56753	&	J.P. Masviel	&	0HP	&	ALPY+Newton200mm	&	429	&	3603 - 7404	&	2739	\\
2014-04-13	&	23:54	&	56761	&	J. Montier	&	C.A.L.C	&	Meade355mm+Atik460E	&	623	&	3841 - 7500	&	3283	\\
2014-04-15	&	21:58	&	56763	&	D. Boyd	&	West\_Challow\_UK	&	C11\_LISA\_SXVR-H694	&	919	&	3900 - 7591	&	4952	\\
2014-04-16	&	22:35	&	56764	&	J. Montier	&	C.A.L.C	&	Meade355mm+Atik460E	&	600	&	3895 - 7578	&	3740	\\
2014-04-17	&	22:35	&	56765	&	C. Buil	&	Castanet	&	C11LORESATIK460EX	&	2080	&	5745 - 6657	&	5418	\\
2014-04-18	&	22:01	&	56766	&	J. Guarro	&	Sta.MariaM.	&	16REMOTATIK460EX	&	732	&	3955 - 7567	&	5355	\\
2014-05-03	&	21:35	&	56781	&	J. Guarro	&	Sta.MariaM.	&	16REMOTATIK460EX	&	650	&	3853 - 7571	&	3019	\\
2014-05-04	&	00:21	&	56781	&	F. Teyssier	&	Rouen	&	SC25cm\_LISA\_ATIK46	&	987	&	4000 - 7551	&	1145	\\
2014-05-04	&	22:48	&	56782	&	J. Montier	&	C.A.L.C	&	Meade355mm+Atik460E	&	636	&	3883 - 7396	&	3513	\\
2014-05-30	&	20:512	&	56808	&	F. Teyssier	&	Rouen	&	SC25cm\_LISA\_ATIK46	&	1010	&	4000 - 7550	&	2152	\\
2014-05-31	&	21:56	&	56809	&	J. Montier	&	C.A.L.C	&	Meade355mm+Atik460E	&	651	&	3914 - 7385	&	3213	\\
2014-06-08	&	03:59	&	56816	&	T. Lester	&	Arnprior\_ON\_CA	&	311DK6.6+23um600+QS	&	1571	&	4011 - 7466	&	3642	\\
2014-06-20	&	21:32	&	56829	&	F. Teyssier	&	Rouen	&	SC25cm\_LISA	&	1030	&	4000 - 7500	&	1844	\\
2014-06-30	&	03:34	&	56838	&	K. Graham	&	GrandLake,CO	&	lx20010"Alpy600A	&	487	&	3965 - 7404	&	1210	\\
2014-08-08	&	22:39	&	56877	&	M. Rodriguez	&	Madrid-Ventilla	&	RC\_0.2m\_f.8\_SC\_Alpy	&	492	&	3765 - 7605	&	3000	\\
2015-02-14	&	23:45	&	57068	&	P. Somogyi	&	Tata	&	25cmf4AlpyAtik414	&	532	&	3651 - 7924	&	2412	\\
2015-02-18	&	02:53	&	57071	&	F. Teyssier	&	Rouen	&	SC25cm\_LISA	&	1066	&	3900 - 7501	&	2426	\\
2015-03-01	&	06:18	&	57082	&	T. Lester	&	MillRidge\_ON\_CA	&	31cmDK+23um1800lpm+	&	8700	&	5992 - 7096	&	6067	\\
2015-03-01	&	06:18	&	57082	&	T. Lester	&	MillRidge\_ON\_CA	&	31cmDK+23um1800lpm+	&	8700	&	5992 - 7096	&	6067	\\
2015-03-20	&	23:36	&	57102	&	P. Somogyi	&	Tata	&	25cmf4AlpyAtik414	&	3723	&	6037 - 6764	&	3606	\\
2015-04-07	&	00:27	&	57119	&	J. Montier	&	C.A.L.C	&	Meade355mm+Atik460E	&	 	&	3838 - 7387	&	1544	\\
2015-04-10	&	23:40	&	57123	&	P. Somogyi	&	Tata	&	25cmLH600Atik414	&	4055	&	6048 - 6766	&	2403	\\
2015-04-13	&	23:38	&	57126	&	J. Montier	&	C.A.L.C	&	Meade355mm+Atik460E	&	 	&	3944 - 7388	&	1544	\\
2015-04-14	&	23:18	&	57127	&	O. Garde	&	OHP	&	C14LISAATIK460EX	&	524	&	3801 - 8071	&	3897	\\
2015-04-16	&	03:36	&	57128	&	T. Lester	&	MillRidge\_ON\_CA	&	31cmDK+23um1800lpm+	&	1971	&	4000 - 7475	&	3633	\\
2015-04-17	&	22:08	&	57130	&	J.Guarro	&	Sta.Mariademontm	&	16REMOTATIK460EX	&	1028	&	3726 - 7448	&	4212	\\

\hline
\end{tabular}
\end{table}
\end{landscape}

\begin{landscape}
\begin{table}
 \centering
\contcaption{}
\label{logspec:continued}
  \begin{tabular}{|ccccccccc|}
  \hline
Date	&	UT	&	MJD	&	Observer	&	Site	&	Instrument	&	Resolution	&	Range [\AA]	&	Exposure [s]	\\
\hline	
2015-04-18	&	22:16	&	57131	&	F. Teyssier	&	Rouen	&	SC14+eshel+460EX	&	11000	&	4211 - 7163	&	3671	\\
2015-04-19	&	22:16	&	57132	&	D. Boyd	&	WestChallowUK	&	C11+LISA+SXVR-H694	&	897	&	3900 - 8100	&	3702	\\
2015-04-20	&	22:00	&	57133	&	P. Somogyi	&	Tata	&	25cmLH24K35u414exm	&	12432	&	6498 - 6611	&	3600	\\
2015-04-27	&	21:06	&	57140	&	F. Teyssier	&	Rouen	&	SC14+eshel+460EX	&	11000	&	4210 - 7167	&	1173	\\
2015-05-08	&	21:24	&	57151	&	F. Teyssier	&	Rouen	&	SC14+eshel+460EX	&	11000	&	4210 - 7167	&	1838	\\
2015-05-08	&	21:24	&	57151	&	F. Teyssier	&	Rouen	&	SC14+eshel+460EX	&	11000	&	4210 - 7166	&	1838	\\
2015-05-11	&	21:34	&	57154	&	P. Somogyi	&	Tata	&	25cmLH24K23u414exm	&	16924	&	6502 - 6614	&	3600	\\
2015-05-11	&	21:33	&	57154	&	P. Somogyi	&	Tata	&	25cmLH24K23u414exm	&	16924	&	6502 - 6614	&	3600	\\
2015-05-16	&	20:52	&	57159	&	F. Teyssier	&	Rouen	&	SC14+eshel+460EX	&	11000	&	4210 - 7167	&	3769	\\
2015-05-16	&	20:52	&	57159	&	F. Teyssier	&	Rouen	&	SC14+eshel+460EX	&	11000	&	4210 - 7166	&	3769	\\
2015-05-18	&	20:06	&	57161	&	P. Berardi	&	BellavistaObs.LA	&	LHIRES31200C9SXV	&	5873	&	6339 - 6789	&	4032	\\
2015-05-22	&	03:14	&	57164	&	K. Graham	&	Manhattan,IL	&	lx20010"AlpyAtik	&	 	&	3710 - 7348	&	2117	\\
2015-05-22	&	03:13	&	57164	&	K. Graham	&	Manhattan,IL	&	lx20010"AlpyAtik	&	 	&	3709 - 7347	&	2117	\\
2015-05-25	&	12:23	&	57168	&	Dong Li	&	JadeTianjin	&	C11LHIRES3-2400\_2x	&	14791	&	6500 - 6611	&	3605	\\
2015-05-25	&	12:23	&	57168	&	Dong Li	&	JadeTianjin	&	C11LHIRES3-2400\_2x	&	14791	&	6500 - 6611	&	3605	\\
2015-05-27	&	21:00	&	57170	&	J. Guarro	&	Sta.Mariademontm	&	16REMOTATIK460EX	&	976	&	3768 - 7390	&	3611	\\
2015-06-06	&	20:19	&	57180	&	P. Somogyi	&	Tata	&	25cmLH24K35u414exm	&	12051	&	6511 - 6623	&	280	\\
2015-07-11	&	21:51	&	57215	&	P. Somogyi	&	Tata	&	25cmLH600\_23u414exm	&	3583	&	6069 - 6791	&	1800	\\
2015-07-11	&	22:34	&	57215	&	P. Somogyi	&	Tata	&	25cmLH600\_23u414exm	&	3090	&	5106 - 5833	&	1800	\\
2015-08-29	&	21:05	&	57264	&	V. Bouttard	&	Osenbach	&	Newton200Alpy600A	&	655	&	3800 - 7800	&	1659	\\
2015-12-31	&	04:27	&	57387	&	P. Somogyi	&	Tata	&	25cmLH600\_23u414exm	&	3331	&	6355 - 7075	&	4903	\\
2016-02-07	&	04:32	&	57425	&	P. Somogyi	&	Tata	&	25cmLH600\_23u414exm	&	2414	&	6385 - 7095	&	3683	\\
2016-02-17	&	03:38	&	57435	&	F. Teyssier	&	Rouen	&	SC14+eShel	&	11000	&	4144 - 7161	&	1843	\\
2016-02-27	&	03:56	&	57445	&	P. Somogyi	&	Tata	&	25cmLH600\_23u414exm	&	2014	&	4608 - 5330	&	600	\\
2016-02-29	&	03:39	&	57447	&	F. Teyssier	&	Rouen	&	SC14+eShel	&	11000	&	4144 - 7161	&	3686	\\
\hline
\end{tabular}
\end{table}
\end{landscape}

\begin{figure*}\centering
\resizebox{\hsize}{!}{\includegraphics{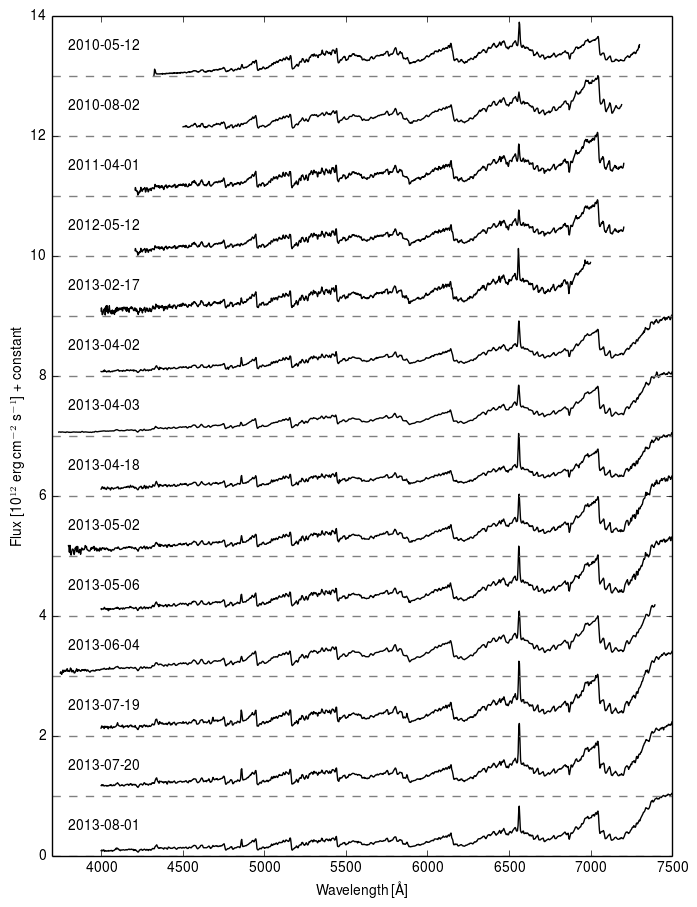}}
  \caption{Spectroscopic observations of T CrB from the ARAS database.}
\label{all_spec}
\end{figure*}

\begin{figure*}\centering
\resizebox{\hsize}{!}{\includegraphics{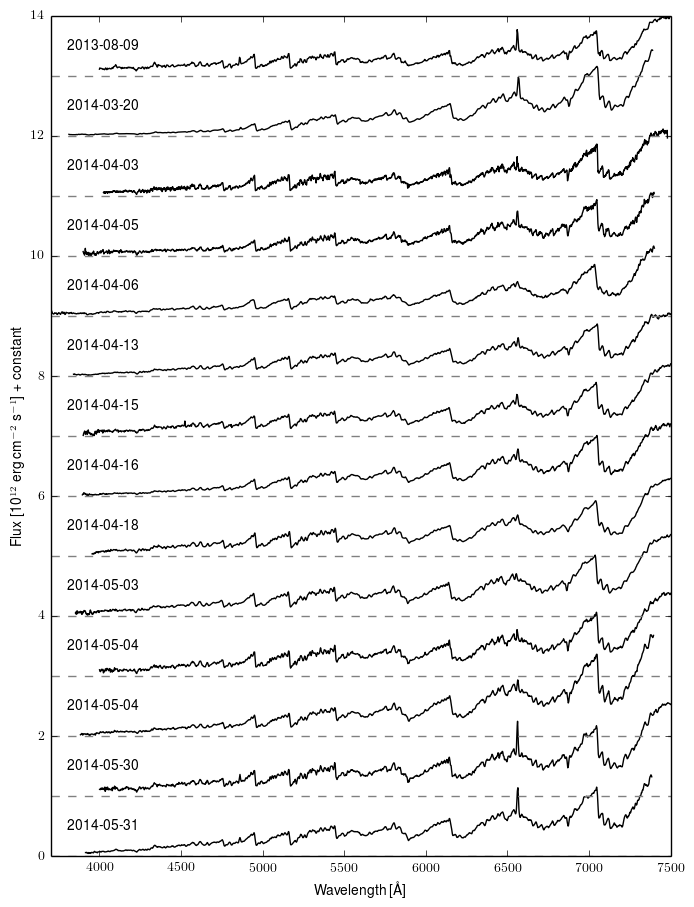}}
\contcaption{}
\label{all_spec_cont}
\end{figure*}

\begin{figure*}\centering
\resizebox{\hsize}{!}{\includegraphics{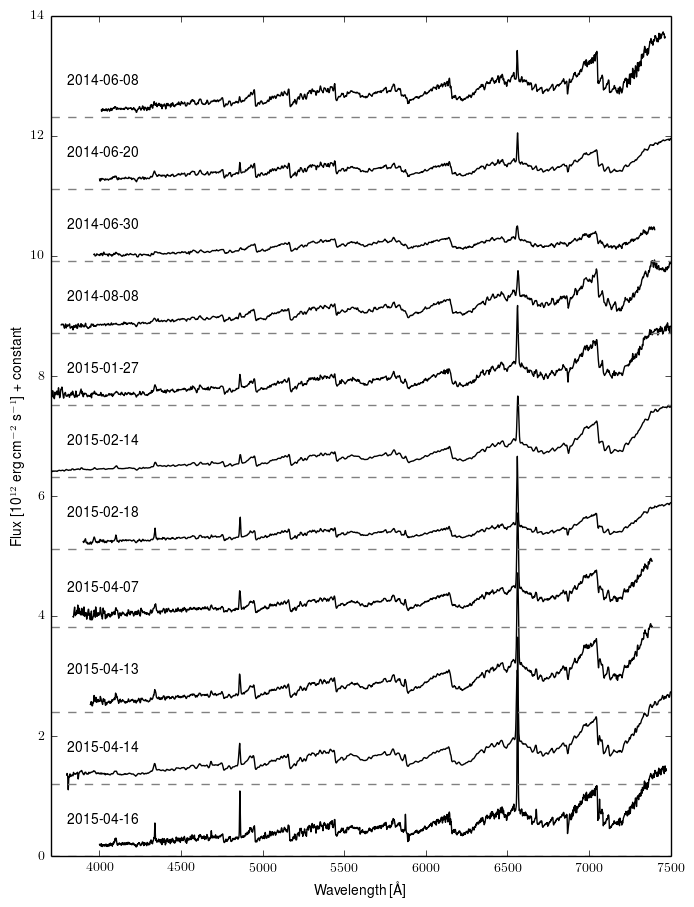}}
\contcaption{}
\label{all_spec_cont}
\end{figure*}

\begin{figure*}\centering
\resizebox{\hsize}{!}{\includegraphics{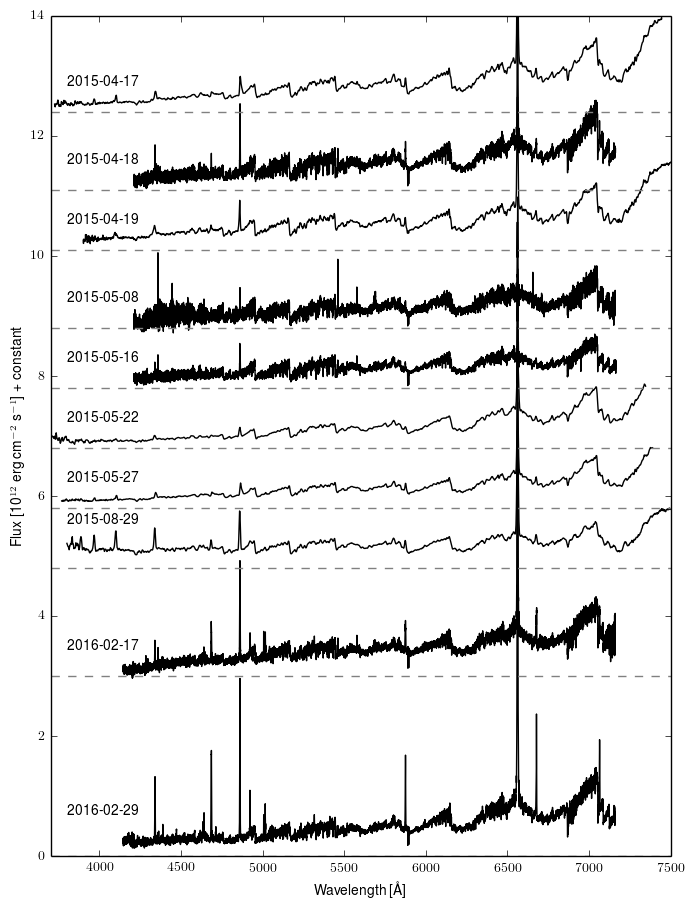}}
\contcaption{}
\label{all_spec_cont}
\end{figure*}

\begin{table*}
 \centering
  \caption{Emission line fluxes in units of 10$^{-13}$ erg cm$^{-2}$ s$^{-1}$. Orbital phase based on ephemeris from  \citet{1988AJ.....95.1505L}. }\label{fluxes_tab}
  \begin{tabular}{|cccccccc|}
  \hline
MJD	&	Date	&	Phase	&	V	&	\mbox{He\,{\sc ii}\,4686}	&	H$\beta$	&	H$\alpha$	&	\mbox{He\,{\sc i}\,5876}	\\
\hline															
55328	&	2010-05-12	&	0.77	&	10.20	&	0.6	&	6	&	41	&	3.2	\\
55410	&	2010-08-02	&	0.13	&	10.17	&		&		&	16	&	2.5	\\
55652	&	2011-04-01	&	0.19	&	10.05	&		&	31	&	25	&		\\
56018	&	2012-04-01	&	0.80	&	9.87	&		&		&	58	&		\\
56059	&	2012-05-12	&	0.98	&	10.19	&		&	3	&	18	&		\\
56340	&	2013-02-17	&	0.21	&	10.06	&	0.3	&	9	&	42	&	3.3	\\
56384	&	2013-04-02	&	0.41	&	10.28	&	1.6	&	11	&	48	&	2.6	\\
56385	&	2013-04-03	&	0.41	&	10.33	&	0.5	&	6	&	40	&		\\
56400	&	2013-04-18	&	0.48	&	10.26	&	1.8	&	11	&	56	&	2.5	\\
56415	&	2013-05-02	&	0.54	&	10.12	&	1.5	&	10	&	51	&		\\
56419	&	2013-05-06	&	0.56	&	10.08	&	0.9	&	13	&	59	&		\\
56448	&	2013-06-04	&	0.69	&	9.98	&	1.4	&	9	&	55	&		\\
56493	&	2013-07-19	&	0.89	&	10.06	&	4.8	&	16	&	68	&	6.3	\\
56494	&	2013-07-20	&	0.89	&	10.06	&	4.2	&	17	&	71	&	6.1	\\
56506	&	2013-08-01	&	0.94	&	10.52	&	1.7	&	10	&	41	&	4.0	\\
56514	&	2013-08-09	&	0.98	&	10.32	&	1.4	&	9.8	&	33	&	3.1	\\
56736	&	2014-03-20	&	0.95	&	10.31	&		&	3.8	&	43	&	1.6	\\
56750	&	2014-04-03	&	0.02	&	10.37	&		&	1.3	&	14	&		\\
56753	&	2014-04-05	&	0.03	&	10.39	&	1.0	&	2.4	&	26	&		\\
56753	&	2014-04-06	&	0.03	&	10.39	&	0.4	&		&	15	&		\\
56761	&	2014-04-13	&	0.06	&	10.29	&		&	1.8	&	17	&		\\
56763	&	2014-04-15	&	0.07	&	10.22	&		&	5.5	&	21	&	1.1	\\
56764	&	2014-04-16	&	0.08	&	10.18	&		&	2.3	&	24	&	0.8	\\
56766	&	2014-04-18	&	0.09	&	10.12	&		&	3.2	&	17	&		\\
56781	&	2014-05-03	&	0.15	&	10.06	&		&		&	11	&		\\
56781	&	2014-05-04	&	0.15	&	10.06	&		&	2.3	&	16	&		\\
56782	&	2014-05-04	&	0.16	&	10.01	&	0.3	&	1.6	&	20	&		\\
56808	&	2014-05-30	&	0.27	&	9.95	&	0.7	&	10	&	52	&		\\
56809	&	2014-05-31	&	0.27	&	9.95	&		&	8.4	&	53	&		\\
56816	&	2014-06-08	&	0.31	&	9.95	&	1.0	&	8.1	&	36	&	3.5	\\
56829	&	2014-06-20	&	0.36	&	10.11	&	2.2	&	13	&	43	&	3.7	\\
56838	&	2014-06-30	&	0.40	&	10.30	&	0.7	&	4.5	&	32	&	2.7	\\
56878	&	2014-08-08	&	0.58	&	9.97	&		&	11	&	51	&		\\
57049	&	2015-01-27	&	0.33	&	9.91	&	1.7	&	20	&	107	&	5.4	\\
57068	&	2015-02-14	&	0.41	&	10.11	&	3.8	&	17	&	94	&	3.5	\\
57071	&	2015-02-18	&	0.43	&	10.36	&	3.3	&	26	&	93	&	5.8	\\
57119	&	2015-04-07	&	0.64	&	9.87	&	6.6	&	30	&	144	&	12.6	\\
57126	&	2015-04-13	&	0.67	&	9.78	&	7.0	&	34	&	179	&	18	\\
57127	&	2015-04-14	&	0.67	&	9.78	&	8.5	&	37	&	152	&	21	\\
57128	&	2015-04-16	&	0.68	&	9.78	&	6.0	&	39	&	173	&	12.0	\\
57130	&	2015-04-17	&	0.68	&	9.78	&	3.7	&	36	&	177	&	12.4	\\
57131	&	2015-04-18	&	0.69	&	9.78	&	3.7	&	34	&	179	&	11.8	\\
57132	&	2015-04-19	&	0.69	&	9.78	&	4.9	&	35	&	148	&	10.0	\\
57140	&	2015-04-27	&	0.73	&	9.99	&		&	8.2	&	281	&	8.2	\\
57151	&	2015-05-08	&	0.78	&	10.14	&		&	11	&	59	&	2.5	\\
57159	&	2015-05-16	&	0.81	&	10.06	&		&	16	&	69	&	3.1	\\
57164	&	2015-05-22	&	0.83	&	10.12	&	3.5	&	17	&	107	&	4.2	\\
57170	&	2015-05-27	&	0.86	&	10.12	&	2.5	&	25	&	120	&	9.4	\\
57264	&	2015-08-29	&	0.27	&	9.95	&	14	&	63	&	196	&	17	\\
57435	&	2016-02-17	&	0.02	&	9.79	&	27	&	59	&	296	&	20	\\
57447	&	2016-02-29	&	0.08	&	9.74	&	45	&	79	&	454	&	34	\\
\hline
\end{tabular}
\end{table*}

\begin{table*}
 \centering
  \caption{Equivalent widths of emission lines. Orbital phase based on ephemeris from  \citet{1988AJ.....95.1505L}. }\label{ew_tab}
  \begin{tabular}{|cccccccc|}
  \hline
MJD	&	Date	&	Phase	&	\mbox{He\,{\sc ii}\,4686}	&	H$\beta$	&	H$\alpha$	&	\mbox{He\,{\sc i}\,5876}	\\
\hline															
55328	&	2010-05-12	&	0.77	&	0.7	&	4.8	&	8.3	&	1.2	\\
55410	&	2010-08-02	&	0.13	&		&		&	2.9	&	0.9	\\
55652	&	2011-04-01	&	0.19	&		&	1.5	&	4.2	&		\\
56018	&	2012-04-01	&	0.80	&		&		&	5.5	&		\\
56059	&	2012-05-12	&	0.98	&		&	1.6	&	5.4	&		\\
56340	&	2013-02-17	&	0.21	&	0.2	&	4.8	&	7.0	&	1.1	\\
56384	&	2013-04-02	&	0.41	&	1.0	&	6.5	&	9.9	&	1.0	\\
56385	&	2013-04-03	&	0.41	&	0.3	&	3.5	&	8.0	&		\\
56400	&	2013-04-18	&	0.48	&	0.9	&	5.7	&	12.3	&	1.0	\\
56415	&	2013-05-02	&	0.54	&	0.7	&	4.7	&	9.0	&		\\
56419	&	2013-05-06	&	0.56	&	0.4	&	5.9	&	10.2	&		\\
56448	&	2013-06-04	&	0.69	&	0.6	&	4.0	&	8.9	&		\\
56493	&	2013-07-19	&	0.89	&	2.0	&	7.0	&	11.6	&	2.2	\\
56494	&	2013-07-20	&	0.89	&	1.6	&	6.7	&	13.1	&	2.2	\\
56506	&	2013-06-01	&	0.94	&	1.2	&	7.0	&	10.2	&	2.2	\\
56514	&	2013-08-09	&	0.98	&	0.8	&	5.2	&	7.7	&	1.5	\\
56736	&	2014-03-20	&	0.95	&		&	3.6	&	6.8	&	0.5	\\
56750	&	2014-04-03	&	0.02	&		&	0.9	&	3.1	&		\\
56753	&	2014-04-05	&	0.03	&	0.9	&	2.0	&	5.2	&		\\
56753	&	2014-04-06	&	0.03	&	0.3	&		&	3.2	&		\\
56761	&	2014-04-13	&	0.06	&		&	1.2	&	3.6	&		\\
56763	&	2014-04-15	&	0.07	&		&	3.2	&	4.1	&	0.4	\\
56764	&	2014-04-16	&	0.08	&		&	1.6	&	4.0	&	0.3	\\
56765	&	2014-04-17	&	0.08	&		&		&	4.7	&	1.0	\\
56766	&	2014-04-18	&	0.09	&		&	1.7	&	3.1	&		\\
56781	&	2014-05-03	&	0.15	&		&		&	1.9	&		\\
56781	&	2014-04-04	&	0.15	&		&	1.2	&	2.7	&		\\
56782	&	2014-05-04	&	0.16	&	0.1	&	0.9	&	2.7	&		\\
56808	&	2014-05-30	&	0.27	&	0.3	&	4.7	&	7.7	&		\\
56809	&	2014-05-31	&	0.27	&		&	3.8	&	7.9	&		\\
56816	&	2014-06-08	&	0.31	&	0.4	&	3.4	&	5.9	&	1.1	\\
56829	&	2014-06-20	&	0.36	&	0.9	&	5.1	&	9.9	&	1.4	\\
56838	&	2014-06-30	&	0.40	&	0.5	&	2.6	&	9.0	&	1.1	\\
56878	&	2014-08-08	&	0.58	&		&	4.3	&	8.3	&		\\
57049	&	2015-01-27	&	0.33	&	0.6	&	7.4	&	16.2	&	1.5	\\
57068	&	2015-02-14	&	0.41	&	1.8	&	7.8	&	16.1	&	1.1	\\
57071	&	2015-02-18	&	0.43	&	1.8	&	13.5	&	24.4	&	2.6	\\
57082	&	2015-03-01	&	0.47	&		&		&	23.0	&		\\
57102	&	2015-03-20	&	0.56	&		&		&	25.9	&		\\
57119	&	2015-04-07	&	0.64	&	2.2	&	10.0	&	22.1	&	3.7	\\
57123	&	2015-04-10	&	0.65	&		&		&	23.3	&		\\
57126	&	2015-04-13	&	0.67	&	2.5	&	12.0	&	22.4	&	4.8	\\
57127	&	2015-04-14	&	0.67	&	2.9	&	11.7	&	21.1	&	6.2	\\
57128	&	2015-04-16	&	0.68	&	2.0	&	13.4	&	22.8	&	2.9	\\
57130	&	2015-04-17	&	0.68	&	1.3	&	12.6	&	21.9	&	3.1	\\
57131	&	2015-05-18	&	0.69	&	1.1	&	11.1	&	21.7	&	2.9	\\
57132	&	2015-04-19	&	0.69	&	1.4	&	11.0	&	21.0	&	2.6	\\
57133	&	2015-04-20	&	0.70	&		&		&	22.3	&		\\
57140	&	2015-04-27	&	0.73	&		&	8.2	&	15.0	&	1.9	\\
57151	&	2015-05-08	&	0.78	&		&	4.7	&	9.6	&	0.8	\\
57154	&	2015-05-11	&	0.79	&		&		&	10.9	&		\\
57159	&	2015-05-16	&	0.81	&		&	7.0	&	13.0	&	1.0	\\
57161	&	2015-05-18	&	0.82	&		&		&	15.0	&		\\
57164	&	2015-05-22	&	0.83	&	1.9	&	9.2	&	16.9	&	1.3	\\
57168	&	2015-05-25	&	0.85	&		&		&	16.7	&		\\
57170	&	2015-05-27	&	0.86	&	1.1	&	11.3	&	21.3	&	3.2	\\
57180	&	2015-06-06	&	0.90	&		&		&	25.7	&		\\
57215	&	2015-07-11	&	0.06	&		&		&	26.1	&		\\
57215	&	2015-07-11	&	0.06	&		&		&		&		\\
57264	&	2015-08-29	&	0.27	&	4.5	&	21.0	&	35.9	&	5.6	\\
57387	&	2015-12-13	&	0.81	&		&		&	36.3	&		\\
57425	&	2016-02-07	&	0.98	&		&		&	42.6	&		\\
57435	&	2016-02-17	&	0.02	&	10.9	&	20.4	&	37.6	&	4.6	\\
57445	&	2016-02-27	&	0.07	&	14.0	&	23.3	&		&		\\
57447	&	2016-02-29	&	0.08	&	16.2	&	26.5	&	48.8	&	7.5	\\
\hline
\end{tabular}
\end{table*}

\begin{figure*}\centering
\includegraphics[width=.33\textwidth]{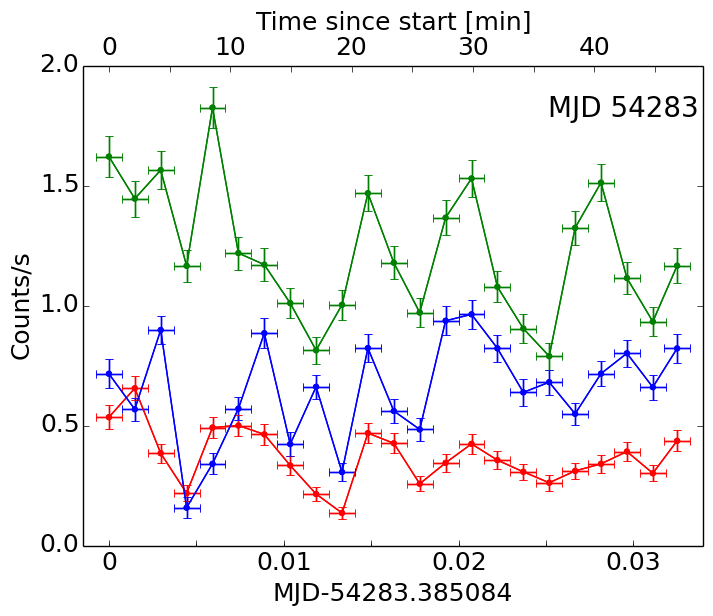}\includegraphics[width=.33\textwidth]{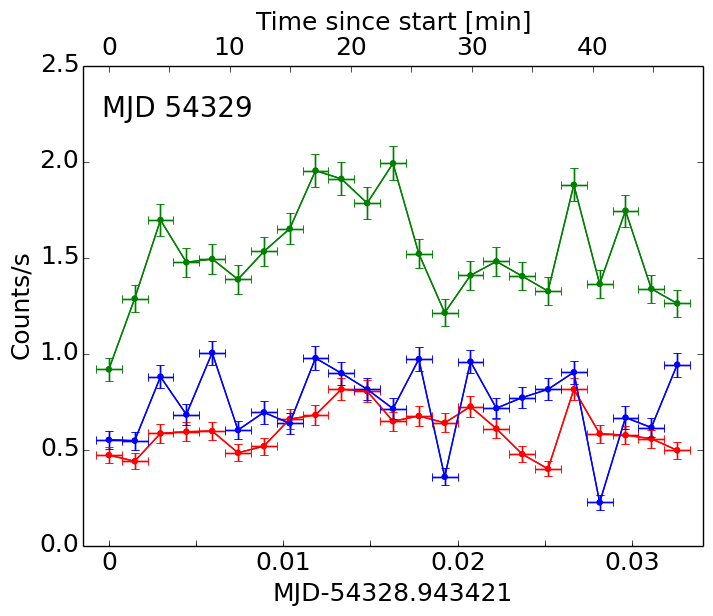}\includegraphics[width=.33\textwidth]{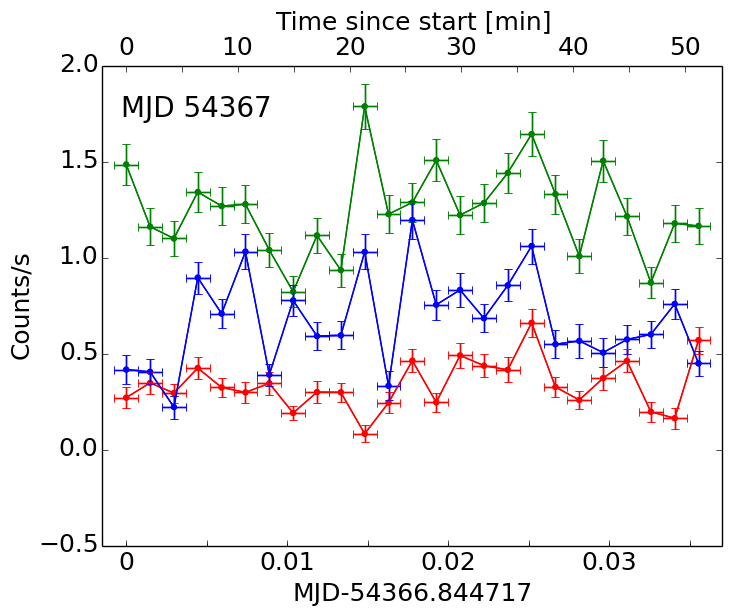}\\
\includegraphics[width=.33\textwidth]{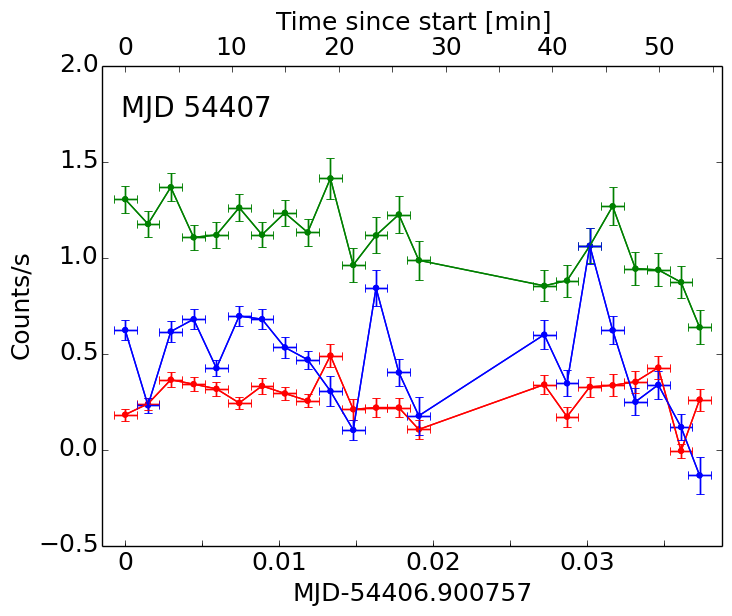}\includegraphics[width=.33\textwidth]{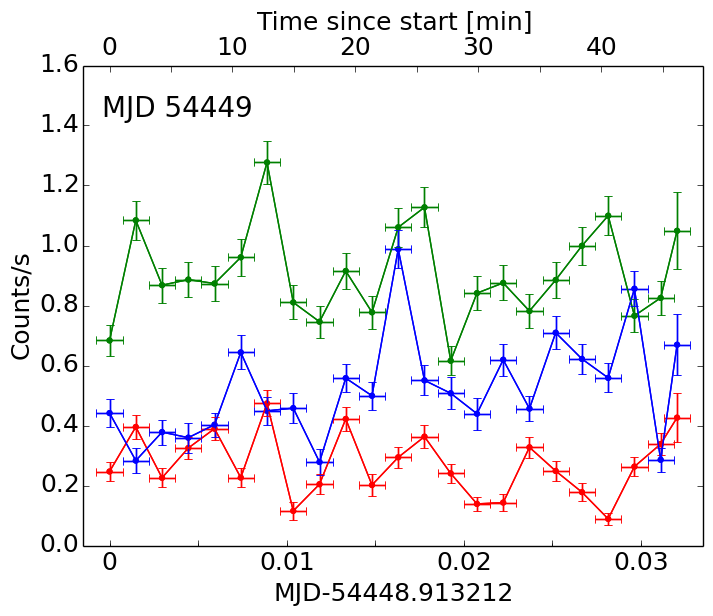}\includegraphics[width=.33\textwidth]{./figs/rxte_flickering/54490}
\includegraphics[width=.33\textwidth]{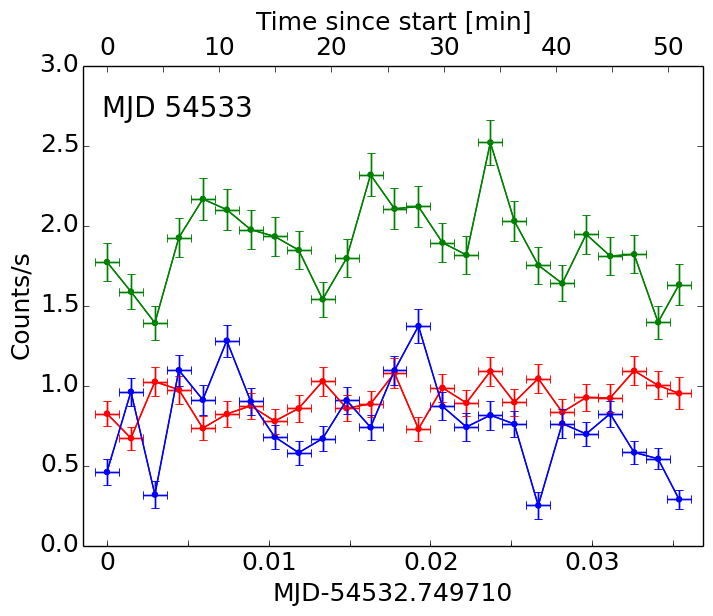}\includegraphics[width=.33\textwidth]{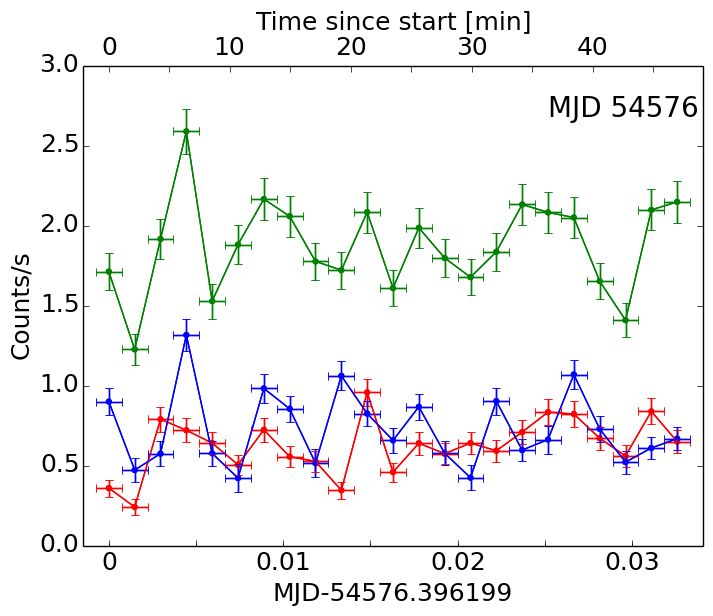}\includegraphics[width=.33\textwidth]{./figs/rxte_flickering/54616}
\includegraphics[width=.33\textwidth]{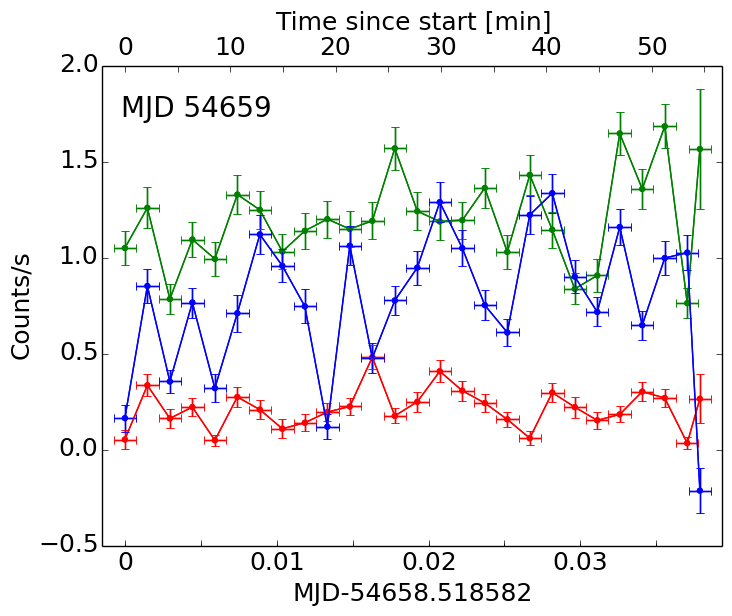}\includegraphics[width=.33\textwidth]{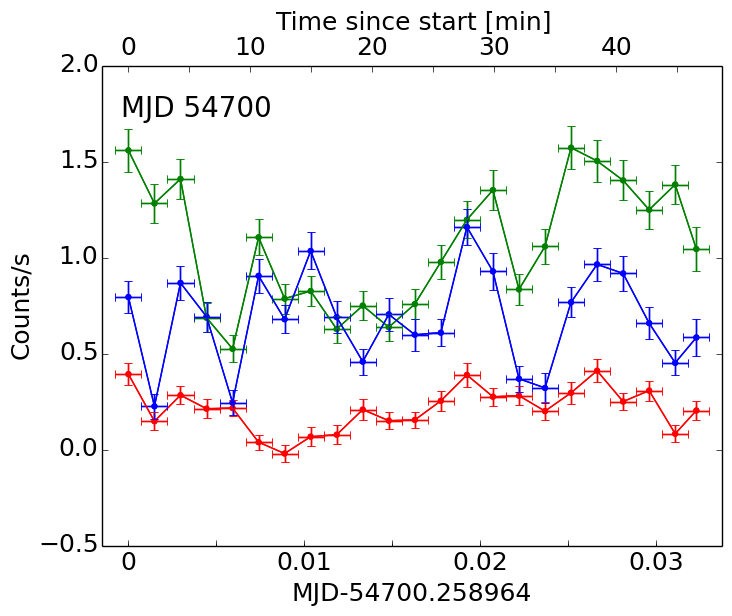}\includegraphics[width=.33\textwidth]{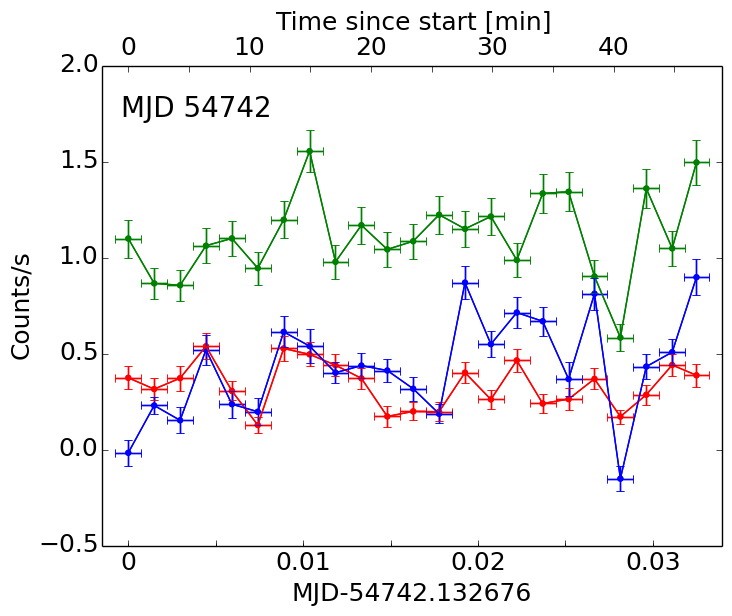}
  \caption{X-ray observations of flickering in T~CrB. The colours are the same as in Fig.~\ref{TCrB_all}.}
\label{rxte_flickering_individual2}
\end{figure*}

\begin{figure*}\centering
\includegraphics[width=.33\textwidth]{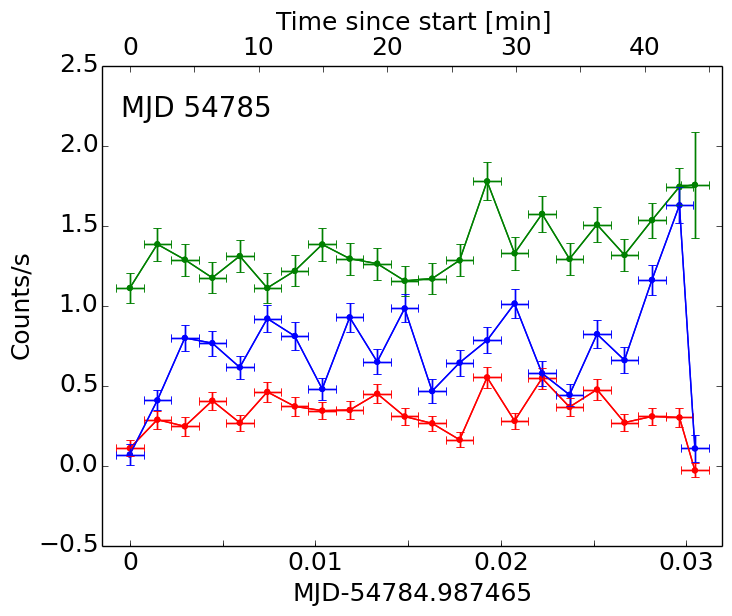}\includegraphics[width=.33\textwidth]{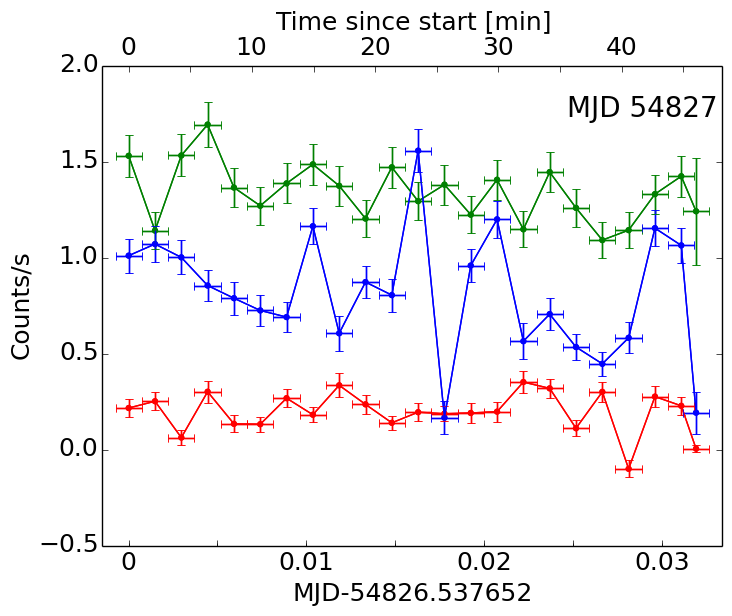}\includegraphics[width=.33\textwidth]{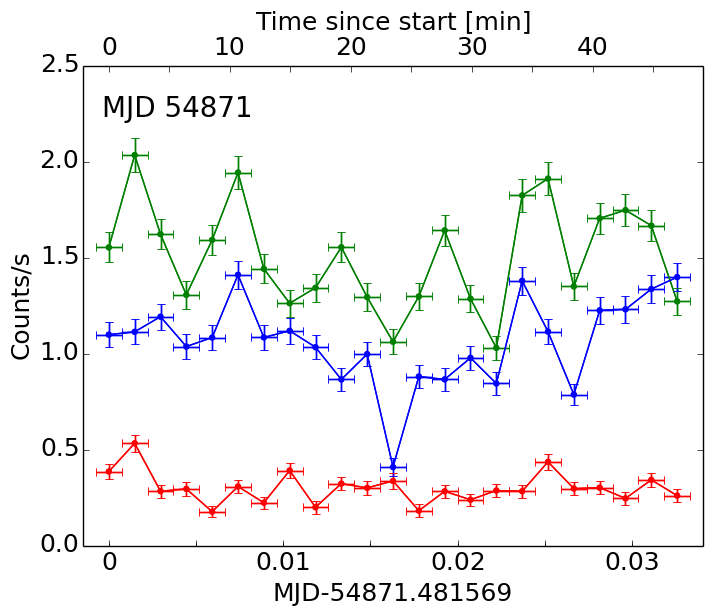}
\includegraphics[width=.33\textwidth]{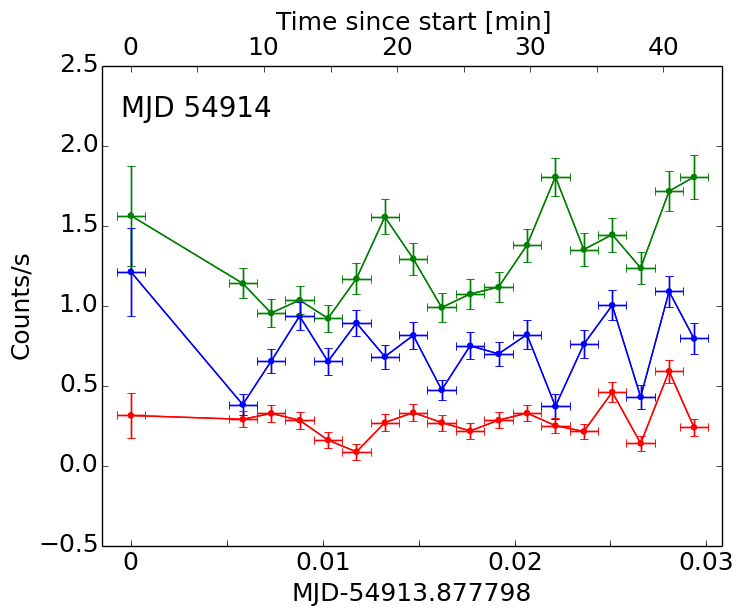}\includegraphics[width=.33\textwidth]{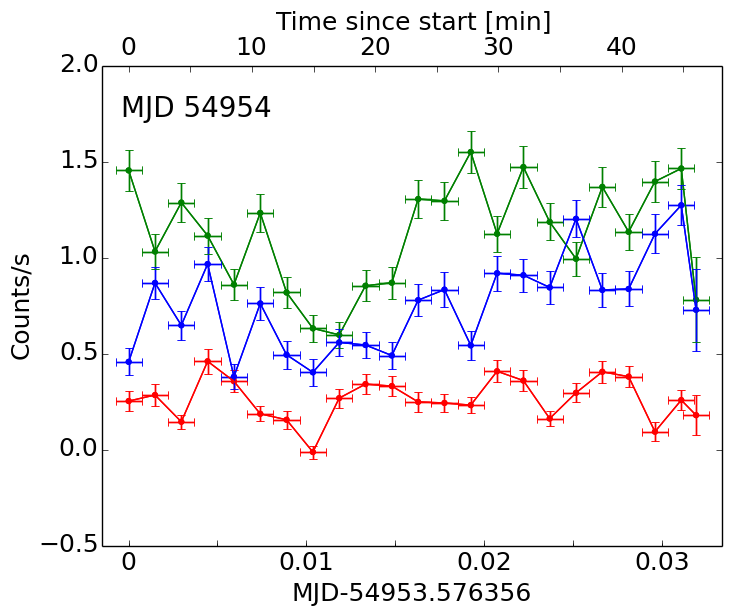}\includegraphics[width=.33\textwidth]{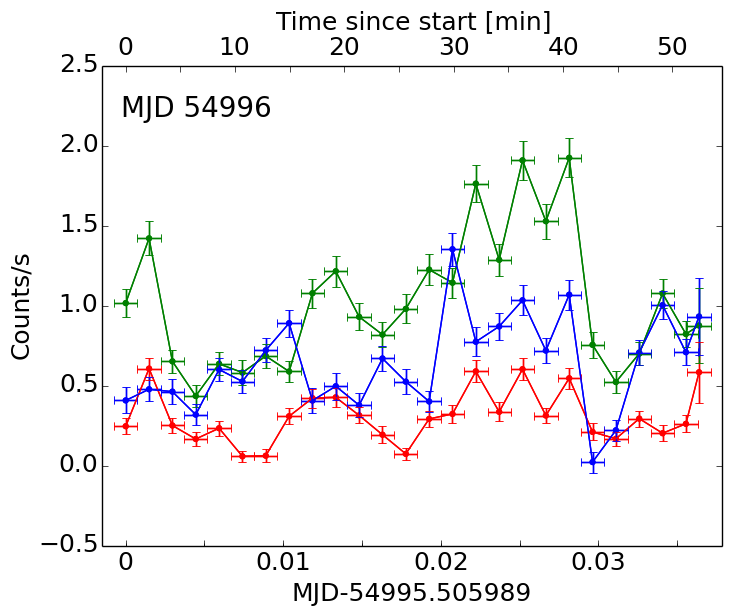}
\includegraphics[width=.33\textwidth]{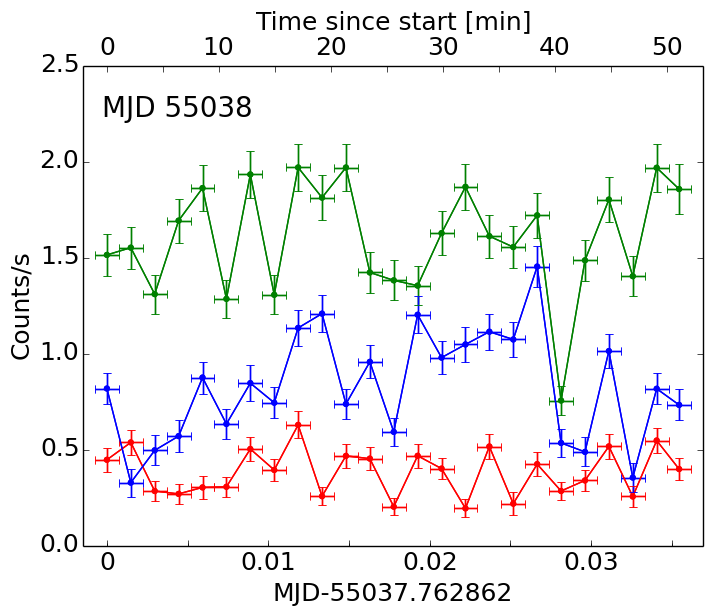}\includegraphics[width=.33\textwidth]{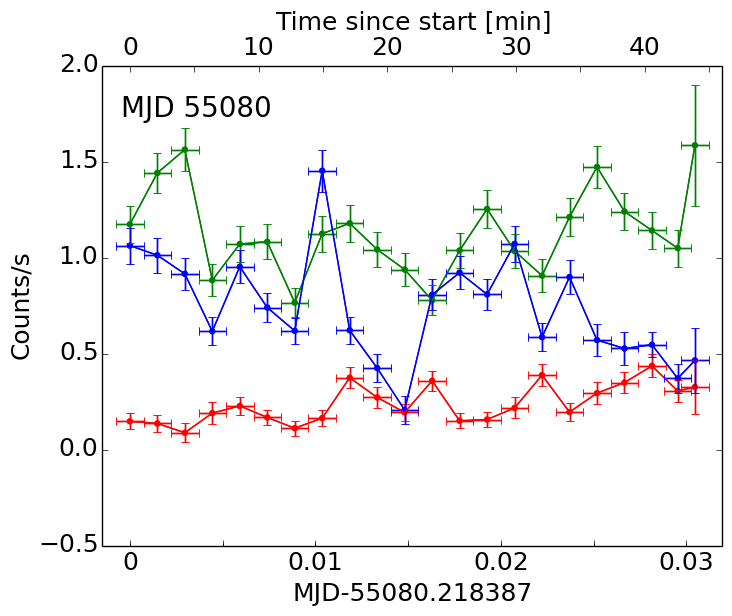}\includegraphics[width=.33\textwidth]{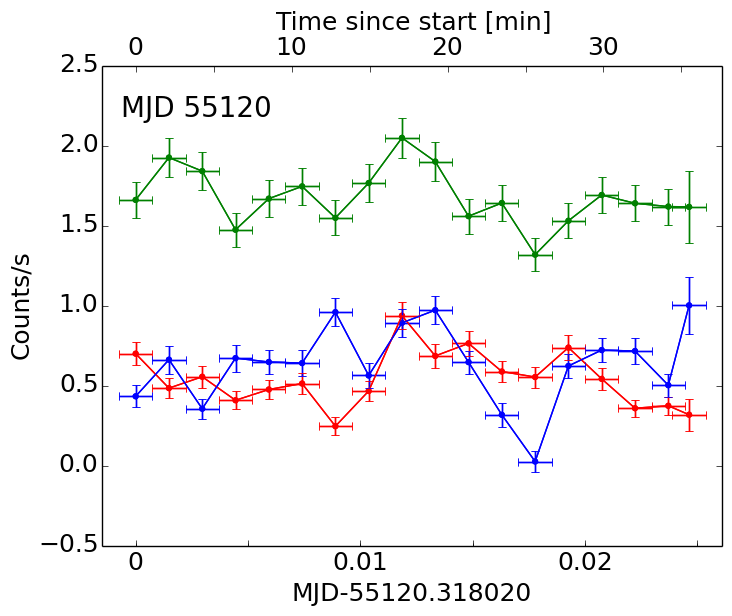}
\includegraphics[width=.33\textwidth]{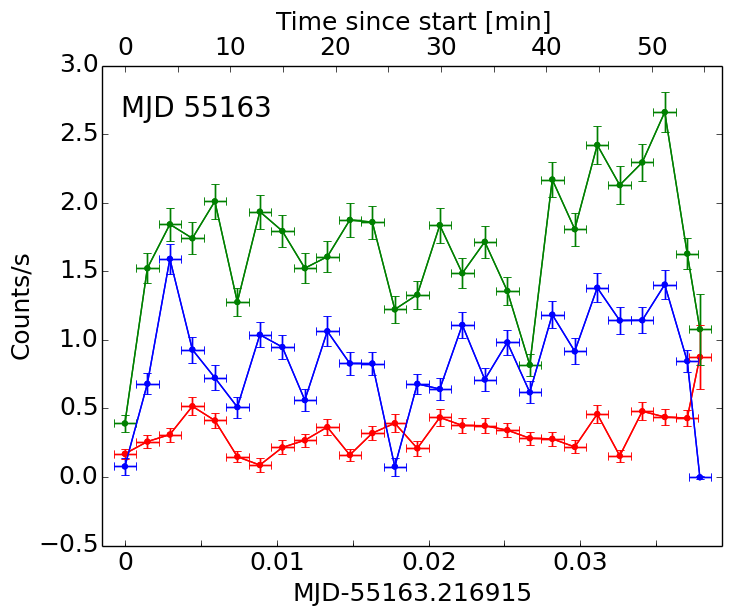}
\contcaption{}
\label{rxte_flickering_individual:continued2}
\end{figure*}

\clearpage
\section{Swift observations}
\label{sec:uvot}

We retrieved publicly available UV images containing T~CrB taken by the Swift UV optical telescope (UVOT; \citealt{2004ApJ...611.1005G}) from the UK Swift Science Data Centre (UKSSDC). \footnote{http://www.swift.ac.uk/index.php} Observations were available in the UV filters UVW2, UVM2 and UVW1, as well as the optical filter $U$. Characteristics of the filters may be found in \citet{2008MNRAS.383..627P}. We followed the UVOT data analysis guide of the UKSSDC to perform aperture photometry on the images using the heasoft-6.18 software\footnote{http://heasarc.gsfc.nasa.gov/lheasoft/} and the latest caldb files for Swift. Optimal aperture sizes dependent on the image filter and pixel binning were used according to \citet{2008MNRAS.383..627P}. Table \ref{tab:swift} gives a log of the observations where the magnitudes are provided in the Vega system. 

\begin{figure}\centering
\resizebox{\hsize}{!}{\includegraphics{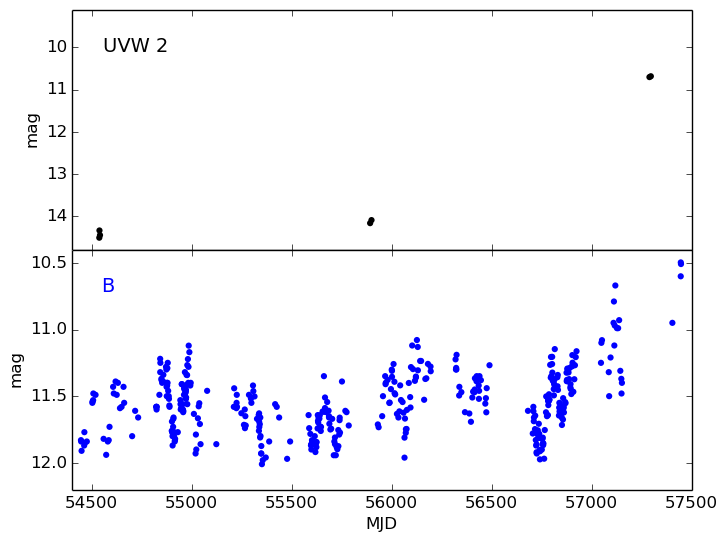}}
  \caption{Optical and UV photometry of T~CrB.}
\label{TCrB_swift}
\end{figure}

The cadence of observations is too low for a meaningful analysis, but it seems that the variability of T~CrB in the UVW2 filter is correlated with variability in the $B$ band (See~Fig.~\ref{TCrB_swift}).

\begin{table}
 \centering
  \caption{Swift UVOT observations of T~CrB. }
  \label{tab:swift}
  \begin{tabular}{|lrrl|}
  \hline
Date	&	Exposure [s]	&	Filter	&	Mag		\\
\hline				
2008-03-13	&	306	&	UVM2	&	14.91$\pm$0.03	\\
2008-03-13	&	265	&	UVW1	&	13.80$\pm$0.02	\\
2008-03-13	&	309	&	UVW2	&	14.52$\pm$0.03	\\
2008-03-14	&	230	&	UVM2	&	15.18$\pm$0.04	\\
2008-03-14	&	177	&	UVW1	&	13.87$\pm$0.02	\\
2008-03-14	&	140	&	UVW2	&	14.35$\pm$0.03	\\
2008-03-16	&	269	&	UVM2	&	14.65$\pm$0.03	\\
2008-03-16	&	255	&	UVW1	&	13.68$\pm$0.02	\\
2008-03-16	&	269	&	UVW2	&	14.46$\pm$0.03	\\
2011-11-27	&	542	&	UVM2	&	14.84$\pm$0.03	\\
2011-11-27	&	465	&	UVW1	&	13.68$\pm$0.02	\\
2011-11-27	&	542	&	UVW2	&	14.17$\pm$0.03	\\
2011-12-04	&	421	&	UVM2	&	14.60$\pm$0.03	\\
2011-12-04	&	363	&	UVW1	&	13.55$\pm$0.02	\\
2011-12-04	&	421	&	UVW2	&	14.10$\pm$0.03	\\
2015-09-23	&	1046	&	U	&	11.81$\pm$0.05	\\
2015-09-24	&	1632	&	UVW2	&	10.71$\pm$0.05	\\
2015-10-01	&	96	   &	U	&	11.81$\pm$0.05	\\
2015-10-01	&	151	&	UVW2	&	10.69$\pm$0.05	\\
\hline
\end{tabular}
\end{table}

\bsp	
\label{lastpage}
\end{document}